# BCS superconductivity in ionic hydrides using chemical capacitor setup

Piotr Szudlarek,[1]‡ Christopher Renskers,[2]‡ Elena R. Margine,[2]* and Wojciech Grochala[1]*

1 Center of New Technologies, University of Warsaw, Zwirki i Wigury 93, 02089 Warsaw Poland; 2 Department of Physics, Applied Physics and Astronomy, Binghamton University, 4400 Vestal Parkway East, Binghamton, NY 13902-6000 USA

*Hydrides, layered materials, phonons, doping, superconductivity*

**ABSTRACT:** We apply a novel "chemical capacitor" setup to facilitate metallization of ionic hydrides, LiH and MgH$_2$. It turns out that the amount of holes doped to a single layer of these materials may reach 0.72 per H atom without structure collapse; concomitant maximum T$_C$ values exceed 17 K in the absence of external pressure for 0.31-hole-doped LiH supported on the LiBaF$_3$ perovskite.

The pursuit of high-temperature superconductors has been a hot topic in condensed matter and materials physics since the discovery of superconductivity (SC) in 1911. The BCS theory of conventional SC predicts that materials based on light elements might exhibit high critical temperatures (T$_C$s). Indeed, the quest for SC in metallic hydrogen (1-4) and in metal hydrides (5-10) has resulted in several hydrogen-rich SC materials with impressive T$_C$ values approaching room temperature. These SC materials encompass H$_3$S (203 K at 155 GPa) (5), LaH$_{x\sim10}$ (260 K at 188 GPa) (6,7), CaH$_6$ (210 K at 160 GPa) (8), and YH$_x$ (262 K at 182 GPa) (9-10). Importantly, some of these materials were predicted by theoretical calculations before they were synthesized (11-14). Other valuable predictions were done based on density functional theory (DFT) (15-18) and many fascinating materials still await synthesis. Unfortunately, all these systems require immense external pressure (p), exceeding 1 mln atm for metallization, and the volumes of the SC phase usually fall below 10$^{-5}$ mm$^3$, rendering their use impractical. Thus, Th$_4$H$_{15}$ (8.3 K), PdH$_{x\sim1}$ (9 K) and (Pd,Ag)H$_{x\sim1}$ (16 K) stand out as record-high T$_C$ metal hydrides at ambient p (19, 20).

Among light-element hydrides, LiH, crystallizing in the NaCl structure with a band gap of nearly 5.0 eV (21), is an immensely stable prototypical ionic hydride that can be melted without thermal decomposition (22). Calculations predict that p-induced metallization and transformation to the CsCl structure occur simultaneously at approximately 330 GPa (23). This p is, therefore, even larger than that needed for metallization of the above-mentioned H-rich materials, and metallization of LiH has never been achieved in a static p experiment.

Recently, we have taken a totally different approach to metallization of broad band gap insulators (24). Nanotechnology allows us to manufacture atomic structures a single-atomic-layer at a time on diverse supports, e.g., via epitaxy, and such systems may readily be modelled by theory. It turns out that the placement of one or more layers of a strong oxidizer in the vicinity of a reductor leads to a substantial charge transfer between the two (Figure 1). For example, if a single layer of AgF$_2$ with a band gap of ca. 3.6 eV (25) is separated by a few layers of inert insulator, RbMgF$_3$, from a layer of LiH with a direct band gap of 5.0 eV, the charge transferred between the two, δ, may reach up to 0.66 electron (e$^-$) per 1 Ag atom or 0.33 hole (h$^+$) per H atom (24). The value of δ is controlled by varying the thickness of the separator between the two. Such a simple setup, called a chemical capacitor, CC, (26), permits *mutual* metallization of both electron-rich and electron-poor layers.

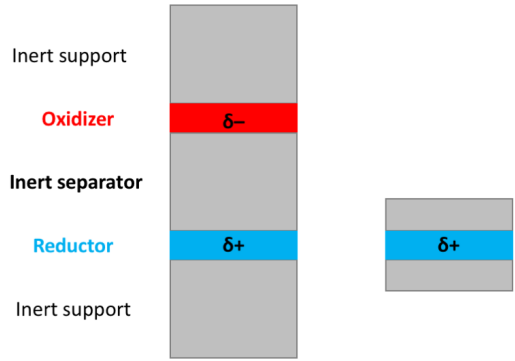

Figure 1. General scheme of a chemical capacitor (left) (25); truncated model used here for calculations of SC in hydrides (right).

Here, we further study this and similar systems, using LiH, NaH or related MgH$_2$ as reductors which may be substantially hole doped in a CC setup. Our purpose is to verify, using DFT calculations, what are the limits of δ in these diverse systems (i.e., the maximum density of holes doped to these insulators, which does not lead to appearance of imaginary phonons), and what T$_C$ values they could reach employing the McMillan-Allen–Dynes formula (27, 28) derived from the Eliashberg theory (29). Since calculations of superconducting properties for systems with more than a few dozen atoms in the unit cell remain prohibitively expensive, we have been forced to truncate the complex setup to merely a single layer of reductor and one or two surrounding layers of support/separator (Figure 1). In the absence of an oxidizer, we simply vary the amount of holes in the hydride layer, and compensate with a uniform jellium background (30). Detailed information on methodology and calculations performed with the Quantum Espresso suite (31) are given in the Supplementary Information (SI).

**Table 1. Tetragonal lattice constant, a, maximum amount of holes per H atom, $\delta_{max}$, maximum DOS at the Fermi level, maximum phonon frequency, $\nu_{max}$, maximum electron-phonon coupling constant, $\lambda_{max}$, and maximum superconducting temperature value, $T_C$, as calculated for over a dozen hydride monolayers placed between inert insulator layers.**

| System | a [Å] | $\delta_{max}$ [h$^+$] | DOS [states Ry$^{-1}$] | $\nu_{max}$ [cm$^{-1}$] | $\lambda_{max}$ [1] | $T_C$ [K] |
|---|---|---|---|---|---|---|
| LiH \| LiF | 3.926 | 0.31 | 3.74 | 898 | 0.86 | 10.7 |
| LiH \| (LiF)$_2$ | 3.975 | 0.29 | 3.65 | 856 | 0.64 | 7.0 |
| LiH \| KMgF$_3$ | 3.997 | 0.25 | 8.3 | 976 | 1.62 | 9.7 |
| LiH \| (BaLiF$_3$) | 3.997 | 0.31 | 7.9 | 935 | 1.20 | 17.0 |
| MgH$_2$ \| RbMgF$_3$ | 3.999 | 0.18 | 6.0 | 1252 | 0.37 | 1.0 |
| LiH \| (BaLiF$_3$)$_2$ | 4.003 | 0.31 | 8.3 | 977 | 1.20 | 17.4 |
| LiH \| (KMgF$_3$)$_2$ | 4.018 | 0.21 | 7.2 | 1003 | 0.37 | 0.8 |
| LiH \| RbMgF$_3$ | 4.051 | 0.25 | 1.2 | 1007 | 0.92 | 14.0 |
| LiH \| (RbMgF$_3$)$_2$ | 4.081 | 0.27 | 10.4 | 991 | 1.13 | 10.5 |
| LiH \| (KZnF$_3$)$_2$ | 4.131 | 0.05 | 3.4 | 1033 | 0.20 | 0 |
| LiH \| KZnF$_3$ | 4.185 | 0.05 | 7.05 | 1403 | 0.21 | 0 |
| NaH \| (CaO)$_2$ | 4.739 | 0.35 | 14.8 | 871 | 0.08 | 0 |
| NaH \| (YN)$_2$ | 4.827 | 0.72 | 19.4 | 819 | 0.18 | 0 |
| NaH \| (LiCl)$_2$ | 4.939 | 0.26 | 13.3 | 585 | 0.15 | 0 |

Results for 14 distinct insulator | hydride | insulator two-dimensional (2D) systems are shown in Table 1; see SI for optimized crystal structures, electronic band structures, phonon spectra, and electron-phonon coupling constants. The systems studied are tetragonal with lattice constant, a, ranging from 3.9 to 4.9 Å. The presence of the lightest element, H, leads to a substantially large maximum frequency that goes up to 1403 cm$^{-1}$ for KZnF$_3$ | LiH | KZnF$_3$. Some systems, such as (KZnF$_3$)$_2$ | LiH | (KZnF$_3$)$_2$, are quite sensitive to doping with phonon instabilities already appearing at 0.1 h$^+$. However, some are quite stable to doping with a maximum hole doping level of 0.72 (for VASP results (36), cf. SI). Their resistance against gap-opening lattice distortions is remarkable, given that doping levels of ⅔, ½, ⅓ or ¼ might suggest facile Peierls distortion (32). For example, a doping level of ⅔ corresponds to a formula H$_2$ + H$^−$ (a polyhydride). While we see evidence of phonon softening as doping level increases, genuine dynamic instability appears in these materials only above $\delta_{max}$ (Figure 2). This is associated with the presence of adjacent support layers sandwiching and straining the hydride monolayer.

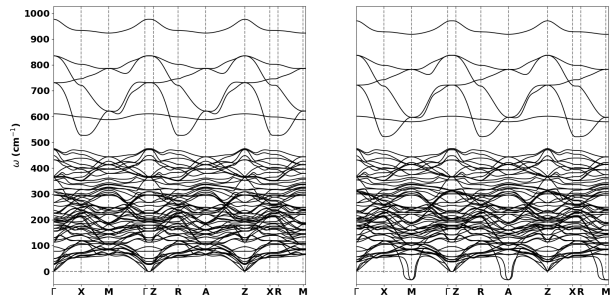

Figure 2. Phonon dispersion for (BaLiF$_3$)$_2$ | LiH | (BaLiF$_3$)$_2$ system for a) $\delta$=0.31; b) $\delta_{max}$=0.32. Note the emergence of a soft and then imaginary phonon at $k_M$=(0.5, 0.5, 0) and $k_A$=(0.5, 0.5, 0.5).

The heavily-doped insulator | hydride | insulator systems are metallic, and the doped band is dominated by contributions from the H (1s) states (Figure 3). These are essentially single-band metals that contain a 2D metallic hydrogen sublattice.

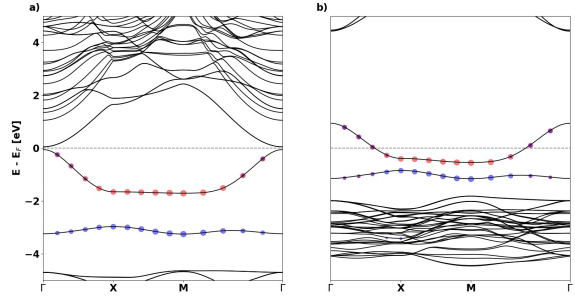

Figure 3. Electronic band structure for (BaLiF$_3$)$_2$ | LiH | (BaLiF$_3$)$_2$ system, a) δ=0; b) δ$_{max}$=0.31. H(1s) character is shown using circles, blue for one H atom, and red for another in the [Li$_2$H$_2$] layer.

In the McMillan–Allen–Dynes expression, the T$_C$ value depends exponentially on the value of electronic density of states (DOS) as well as on the strength of the electron-phonon coupling, λ (cf. SI for detailed formula). In our simple one-band systems DOS increases monotonically with δ (cf. SI); therefore, maximum DOS is reached at δ$_{max}$. The increase in DOS also causes a phonon softening, leading, in turn, to an increase in λ (32) (Figure 4, cf. SI). Table 1 lists the resulting T$_C$ values.

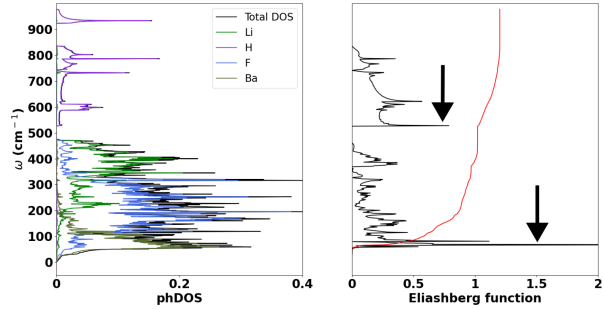

Figure 4. Phonon DOS and isotropic Eliashberg function computed for (BaLiF$_3$)$_2$ | LiH | (BaLiF$_3$)$_2$ system for δ$_{max}$=0.31.

It turns out that the systems studied exhibit electronic total DOS values up to 19.4 states Ry$^{-1}$ for (YN)$_2$ | NaH | (YN)$_2$. This value is comparable to that for "usual" metals, such as Li (DOS of 6.3 states Ry$^{-1}$ is computed for bcc bulk structure). On the other hand, λ$_{max}$ ranges from 0.07 to 1.62; the upper limit corresponds to strong coupling (λ>1) and exceeds that for metallic lead (λ=1.55) (33). Consequently, the computed T$_C$ value, for systems where both DOS and λ are large, exceeds 10 K, with a maximum of 17.4 K for (BaLiF$_3$)$_2$ | LiH | (BaLiF$_3$)$_2$. This value is slightly larger than that of 16 K found experimentally for (Pd,Ag)H$_{x\sim1}$ (20).

Next, it is instructive to gain insight into the character of the phonons that couple strongly with the electrons in the systems studied (Figure 5). It is natural to expect that phonons connected with motions of H atoms will strongly couple to electrons in what is essentially metallic hydrogen. Indeed, a 527 cm$^{-1}$ mode (value at Γ) involving mostly in-plane motion of H atoms substantially contributes to the electron-phonon coupling. Also, it turns out that, some low-frequency phonons couple strongly with electrons. One good example is that of the 70 cm$^{-1}$ phonon for the (BaLiF$_3$)$_2$ | LiH | (BaLiF$_3$)$_2$ system. This mode involves motion of heavy atoms (Ba) while coupled with some out-of-plane contribution from H. Similar features were seen for systems featuring RbMgF$_3$ sandwich layers, with substantial contribution to λ from Rb-dominated phonons. This result shows that seemingly inert sandwiching layers may strongly influence SC in the hydride layer via proximity effects (34, 35).

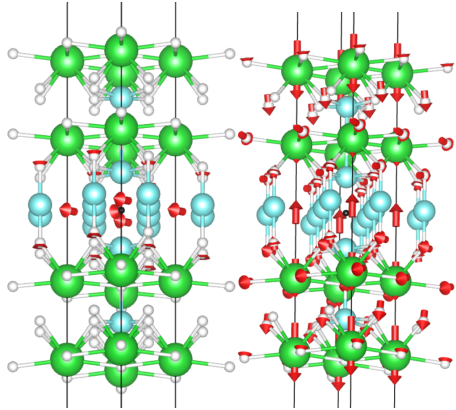

Figure 5. Atomic displacements for two selected phonons which show appreciable electron-phonon coupling for (BaLiF$_3$)$_2$ | LiH | (BaLiF$_3$)$_2$ system for δ$_{max}$=0.31: 527 cm$^{-1}$ (left) and 70 cm$^{-1}$ (right) (as highlighted also in Fig.4 by black arrows).

Stoichiometries and structures studied here are simple, while the Periodic Table of chemical elements offers an immense playground for further explorations. The results presented here encourage us to further pursue theoretical quest for superconducting H-rich systems with even higher $T_C$ values at ambient p conditions.

## ASSOCIATED CONTENT

### Supporting Information

The Supporting Information is available free of charge on the ACS Publications website. SI (PDF) contains optimized crystal structures, electronic band structure, electronic DOS, phonon band structure, phonon DOS, and electron-phonon coupling as well as selected results obtained with VASP (36).

## AUTHOR INFORMATION

### Corresponding Authors

* rmargine@binghamton.edu, w.grochala@cent.uw.edu.pl.

### Author Contributions

The manuscript was written through contributions of all authors. All authors have given approval to the final version of the manuscript. ‡These authors contributed equally.

### Funding Sources

P.S. and W.G. acknowledge support from the Polish National Science Center via project 2021/41/B/ST5/00195, as well as from The University of Warsaw (New Ideas POB II). C.R. and E.R.M. acknowledge support from the National Science Foundation (NSF) under Grant No. DMR-2035518. This work used the supercomputer at ICM UW (project GA83-34). This work also used the Expanse system at the San Diego Supercomputer Center through allocation TG-DMR180071 from the Advanced Cyberinfrastructure Coordination Ecosystem: Services & Support (ACCESS) program [37], which is supported by National Science Foundation grants #2138259, #2138286, #2138307, #2137603, and #2138296.

## ACKNOWLEDGMENT

Dr. Adam Grzelak is thanked for performing initial screening for the LiH | LiF system. Prof. Jose Lorenzana (CNR, Rome) is thanked for insightful remarks to this work.

## ABBREVIATIONS

CC, chemical capacitor; BCS, Bardeen-Cooper-Schrieffer model; SC, superconductivity; 2D, two-dimensional; $T_C$, superconducting critical temperature; DFT, density functional theory; CPU, central processing unit; DOS, density of states; p, pressure.

SYNOPSIS TOC. Single-layer ionic hydrides sandwiched between two support layers may resist hole doping up to substantial doping levels, as follows from the density functional theory study. The resulting two-dimensional metallic hydrides (Fermi surface is shown) are calculated to exhibit superconductivity with $T_C$ values up to 17.4 K at ambient pressure conditions.

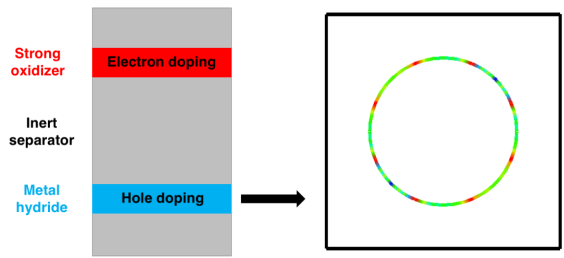

**Supplementary Information**

**BCS superconductivity in ionic hydrides using chemical capacitor setup**
Piotr Szudlarek, Christopher Renskers, Elena R. Margine, and Wojciech Grochala

**Contents**
**S1. Methodology of DFT calculations.**
**S2. Crystallographic Information Files for systems studied, at null and maximum doping levels.**
**S3. Electronic structure of systems studied at null and maximum doping levels.**
**S4. Phonon dispersion curves for systems studied at null and maximum doping levels.**
**S5. Dependence of DOS on a doping level as exemplified by $RbMgF_3$ | LiH | $RbMgF_3$ system.**
**S6. Dependence of λ on a doping level as exemplified by $RbMgF_3$ | LiH | $RbMgF_3$ system.**
**S7. Dependence of $T_C$ on a doping level as exemplified by $RbMgF_3$ | LiH | $RbMgF_3$ system.**
**S8. Dependence of average logarithmic frequency on doping level as exemplified by $RbMgF_3$ | LiH | $RbMgF_3$ system.**
**S9. Dependence of $T_C$ on broadening parameter as exemplified by LiF | LiH | LiF system at 31% doping.**
**S10. Comparison of maximum doping level as computed using QE and VASP for select systems.**

**S1. Methodology of DFT calculations.**
We first built bulk 3D models of the novel "chemical capacitor" setup consisting of the inert support, inert separator, oxidizer, and reductor as shown on the left side in Figure 1 of the main text. Due to the computational complexity of *ab-initio* calculations of such systems, in particular superconductivity, we considered a truncated 2D model that consists only of the inert support, inert separator, and reductor as shown on the right side in Figure 1 of the main text. Next, first-principles calculations were carried out with the Quantum ESPRESSO (QE) package [1,2]. We used the Perdew-Burke-Ernzerhof (PBE) [3] generalized gradient approximation and optimized norm-conserving Vanderbilt (ONCV) pseudopotentials from the Pseudo Dojo library [4]. A plane wave kinetic energy cutoff of 100 Ry for the wavefunctions and 400 Ry for the charge density and potential were used for all systems. A large cutoff of 100 Ry was needed to converge the energy for the RbMgF3-LiH system and was subsequently kept for all other systems for uniformity and ease. For the Brillouin-zone integration, we employed a Γ-centered 12×12×1 **k**-mesh [5] with a Methfessel-Paxton smearing width [6] of 0.01 Ry. The lattice parameters for the undoped systems are all relatively the same justifying the use of a 12×12×1 **k**-mesh for all systems except for the LiF-LiH and CaO-NaH systems which have smaller lattice parameters. For the latter, we used slightly denser **k**-mesh of 18×18×1 and 20×20×1, respectively. Since these are 2D materials, we only sampled with one point in the out of plane direction, and used a vacuum distance of at least 12 Å for all systems. The atomic positions and in-plane lattice parameters were optimized until the total energy was converged within $10^{-6}$ Ry and the force on each atom was less than $10^{-4}$ Ry/Å while keeping the vacuum spacing constant.

To study hole doped systems, we employed a jellium model and relaxed the atomic positions while holding the lattice parameters fixed at the values found for the corresponding undoped systems. The systems were hole doped until they became dynamically unstable (i.e., the appearance of imaginary phonons). A doping level as high as 0.72 electrons per H was reached in NaH | (YN)$_2$. The dynamical matrices and electron-phonon (e-ph) matrix elements were computed using density-functional perturbation theory (DFPT) [7]. We used a 12×12×1 **k**-mesh and 4×4×1 **q**-mesh for all systems, except for LiF-LiH and CaO-NaH for which we used an 18×18×1 **k**-mesh and 6×6×1 **q**-mesh and a 20×20×1 **k**-mesh and 10×10×1 **q**-mesh, respectively. For each **q**-point, the e-ph matrix elements were linearly interpolated to a denser **k**-mesh [8]. The denser mesh was 60×60×1 **k**-mesh for all systems, except for CaO-NaH where an 80×80×1 **k**-mesh was used.

Finally, the Eliashberg spectral function ($\alpha^2 F$) was computed for a set of broadenings from 0 to 0.08 Ry by interpolating the phonons to a finer 60×60×1 **q**-mesh using the tetrahedron method [8]. We identified 0.03 Ry as the optimal choice for smearing and estimated the superconducting critical temperature ($T_c$) using the Allen-Dynes modified McMillan formula [9, 10] with a value of $\mu^* = 0.10$.

The equations for the relevant superconducting quantities: Eliashberg spectral function ($\alpha^2 F$), e-ph coupling strength ($\lambda$), omega log ($\omega_{\log}$), and superconducting critical temperature ($T_c$) are shown below.

$$\alpha^2 F(\omega) = \frac{1}{N_F} \sum_{n\mathbf{k},m\mathbf{q},\nu} |g_\nu(n\mathbf{k}, m\mathbf{q})|^2 \delta(\epsilon_{n\mathbf{k}} - \epsilon_F) \delta(\epsilon_{m\mathbf{k}+\mathbf{q}} - \epsilon_F) \delta(\hbar\omega - \hbar\omega_{\mathbf{q}\nu}) \quad (1)$$

$$\lambda = 2 \int_0^\infty \frac{\alpha^2 F(\omega)}{\omega} d\omega \quad (2)$$

$$\omega_{\log} = \exp\left[\frac{2}{\lambda} \int_0^\infty d\omega \frac{\alpha^2 F(\omega)}{\omega} \log \omega \right] \quad (3)$$

$$k_B T_c = \frac{\hbar \omega_{\log}}{1.2} \exp\left[-\frac{1.04(1+\lambda)}{\lambda - \mu^*(1 + 0.62\lambda)}\right] \quad (4)$$

**S2. Crystallographic Information Files for systems studied, at null and maximum doping levels.**

LiF | LiH | LiF – 0%

```
# CIF file created by FINDSYM, version 7.1.3

data_findsym-output
_audit_creation_method FINDSYM

_cell_length_a     2.7729000000
_cell_length_b     2.7729000000
_cell_length_c    18.4964000000
_cell_angle_alpha 90.0000000000
_cell_angle_beta  90.0000000000
_cell_angle_gamma 90.0000000000
_cell_volume     142.2183462771

_symmetry_space_group_name_H-M "P 4/m 2/m 2/m"
_symmetry_Int_Tables_number 123
_space_group.reference_setting '123:-P 4 2'
_space_group.transform_Pp_abc a,b,c;0,0,0
```

```
loop_
_space_group_symop_id
_space_group_symop_operation_xyz
1 x,y,z
2 x,-y,-z
3 -x,y,-z
4 -x,-y,z
5 -y,-x,-z
6 -y,x,z
7 y,-x,z
8 y,x,-z
9 -x,-y,-z
10 -x,y,z
11 x,-y,z
12 x,y,-z
13 y,x,z
14 y,-x,-z
15 -y,x,-z
16 -y,-x,z

loop_
_atom_site_label
_atom_site_type_symbol
_atom_site_symmetry_multiplicity
_atom_site_Wyckoff_symbol
_atom_site_fract_x
_atom_site_fract_y
_atom_site_fract_z
_atom_site_occupancy
_atom_site_fract_symmform
Li1 Li   2 g   0.00000   0.00000   0.39077   1.00000 0,0,Dz
Li2 Li   1 d   0.50000   0.50000   0.50000   1.00000 0,0,0
F1  F    2 h   0.50000   0.50000   0.38766   1.00000 0,0,Dz
H1  H    1 b   0.00000   0.00000   0.50000   1.00000 0,0,0

# end of cif
```

| LiF \| LiH \| LiF – 31% |
|---|

```
# CIF file created by FINDSYM, version 7.1.3

data_findsym-output
_audit_creation_method FINDSYM

_cell_length_a    2.7762000000
_cell_length_b    2.7762000000
_cell_length_c    18.4964000000
_cell_angle_alpha 90.0000000000
_cell_angle_beta  90.0000000000
_cell_angle_gamma 90.0000000000
_cell_volume      142.5570529088
```

_symmetry_space_group_name_H-M "P 4/m 2/m 2/m"
_symmetry_Int_Tables_number 123
_space_group.reference_setting '123:-P 4 2'
_space_group.transform_Pp_abc a,b,c;0,0,0

loop_
_space_group_symop_id
_space_group_symop_operation_xyz
1 x,y,z
2 x,-y,-z
3 -x,y,-z
4 -x,-y,z
5 -y,-x,-z
6 -y,x,z
7 y,-x,z
8 y,x,-z
9 -x,-y,-z
10 -x,y,z
11 x,-y,z
12 x,y,-z
13 y,x,z
14 y,-x,-z
15 -y,x,-z
16 -y,-x,z

loop_
_atom_site_label
_atom_site_type_symbol
_atom_site_symmetry_multiplicity
_atom_site_Wyckoff_symbol
_atom_site_fract_x
_atom_site_fract_y
_atom_site_fract_z
_atom_site_occupancy
_atom_site_fract_symmform
Li1 Li  2 g  0.00000  0.00000  0.37431  1.00000 0,0,Dz
Li2 Li  1 d  0.50000  0.50000  0.50000  1.00000 0,0,0
F1  F   2 h  0.50000  0.50000  0.38809  1.00000 0,0,Dz
H1  H   1 b  0.00000  0.00000  0.50000  1.00000 0,0,0

# end of cif

LiF | LiF | LiH | LiF | LiF – 0%

# CIF file created by FINDSYM, version 7.1.3

data_findsym-output
_audit_creation_method FINDSYM

_cell_length_a     2.8105000000
_cell_length_b     2.8105000000

```
_cell_length_c    21.9721000000
_cell_angle_alpha 90.0000000000
_cell_angle_beta  90.0000000000
_cell_angle_gamma 90.0000000000
_cell_volume      173.5556459040

_symmetry_space_group_name_H-M "P 4/m 2/m 2/m"
_symmetry_Int_Tables_number 123
_space_group.reference_setting '123:-P 4 2'
_space_group.transform_Pp_abc a,b,c;0,0,0

loop_
_space_group_symop_id
_space_group_symop_operation_xyz
1 x,y,z
2 x,-y,-z
3 -x,y,-z
4 -x,-y,z
5 -y,-x,-z
6 -y,x,z
7 y,-x,z
8 y,x,-z
9 -x,-y,-z
10 -x,y,z
11 x,-y,z
12 x,y,-z
13 y,x,z
14 y,-x,-z
15 -y,x,-z
16 -y,-x,z

loop_
_atom_site_label
_atom_site_type_symbol
_atom_site_symmetry_multiplicity
_atom_site_Wyckoff_symbol
_atom_site_fract_x
_atom_site_fract_y
_atom_site_fract_z
_atom_site_occupancy
_atom_site_fract_symmform
Li1 Li  2 h  0.50000  0.50000  0.68550  1.00000 0,0,Dz
Li2 Li  2 g  0.00000  0.00000  0.59388  1.00000 0,0,Dz
Li3 Li  1 d  0.50000  0.50000  0.50000  1.00000 0,0,0
F1  F   2 g  0.00000  0.00000  0.68849  1.00000 0,0,Dz
F2  F   2 h  0.50000  0.50000  0.59338  1.00000 0,0,Dz
H1  H   1 b  0.00000  0.00000  0.50000  1.00000 0,0,0

# end of cif
```

LiF | LiF | LiH | LiF | LiF – 29%

```
# CIF file created by FINDSYM, version 7.1.3

data_findsym-output
_audit_creation_method FINDSYM

_cell_length_a    2.8104000000
_cell_length_b    2.8104000000
_cell_length_c    21.9721000000
_cell_angle_alpha 90.0000000000
_cell_angle_beta  90.0000000000
_cell_angle_gamma 90.0000000000
_cell_volume      173.5432956063

_symmetry_space_group_name_H-M "P 4/m 2/m 2/m"
_symmetry_Int_Tables_number 123
_space_group.reference_setting '123:-P 4 2'
_space_group.transform_Pp_abc a,b,c;0,0,0

loop_
_space_group_symop_id
_space_group_symop_operation_xyz
1 x,y,z
2 x,-y,-z
3 -x,y,-z
4 -x,-y,z
5 -y,-x,-z
6 -y,x,z
7 y,-x,z
8 y,x,-z
9 -x,-y,-z
10 -x,y,z
11 x,-y,z
12 x,y,-z
13 y,x,z
14 y,-x,-z
15 -y,x,-z
16 -y,-x,z

loop_
_atom_site_label
_atom_site_type_symbol
_atom_site_symmetry_multiplicity
_atom_site_Wyckoff_symbol
_atom_site_fract_x
_atom_site_fract_y
_atom_site_fract_z
_atom_site_occupancy
_atom_site_fract_symmform
Li1 Li  2 h  0.50000  0.50000  0.69933  1.00000 0,0,Dz
Li2 Li  2 g  0.00000  0.00000  0.60399  1.00000 0,0,Dz
```

| |
|---|
| Li3  Li   1 d  0.50000  0.50000  0.50000  1.00000 0,0,0<br>F1   F    2 g  0.00000  0.00000  0.69303  1.00000 0,0,Dz<br>F2   F    2 h  0.50000  0.50000  0.59252  1.00000 0,0,Dz<br>H1   H    1 b  0.00000  0.00000  0.50000  1.00000 0,0,0<br><br># end of cif |
| KMgF$_3$ \| LiH \| KMgF$_3$ – 0% |
| # CIF file created by FINDSYM, version 7.1.3<br><br>data_findsym-output<br>_audit_creation_method FINDSYM<br><br>_cell_length_a     3.9972000000<br>_cell_length_b     3.9972000000<br>_cell_length_c     22.2224000000<br>_cell_angle_alpha  90.0000000000<br>_cell_angle_beta   90.0000000000<br>_cell_angle_gamma  90.0000000000<br>_cell_volume       355.0607924636<br><br>_symmetry_space_group_name_H-M "P 4/m 2/m 2/m"<br>_symmetry_Int_Tables_number 123<br>_space_group.reference_setting '123:-P 4 2'<br>_space_group.transform_Pp_abc a,b,c;0,0,0<br><br>loop_<br>_space_group_symop_id<br>_space_group_symop_operation_xyz<br>1 x,y,z<br>2 x,-y,-z<br>3 -x,y,-z<br>4 -x,-y,z<br>5 -y,-x,-z<br>6 -y,x,z<br>7 y,-x,z<br>8 y,x,-z<br>9 -x,-y,-z<br>10 -x,y,z<br>11 x,-y,z<br>12 x,y,-z<br>13 y,x,z<br>14 y,-x,-z<br>15 -y,x,-z<br>16 -y,-x,z<br><br>loop_<br>_atom_site_label<br>_atom_site_type_symbol<br>_atom_site_symmetry_multiplicity<br>_atom_site_Wyckoff_symbol |

_atom_site_fract_x
_atom_site_fract_y
_atom_site_fract_z
_atom_site_occupancy
_atom_site_fract_symmform
K1  K   2 h  0.50000  0.50000  0.31877  1.00000 0,0,Dz
Mg1 Mg  2 g  0.00000  0.00000  0.40217  1.00000 0,0,Dz
Li1 Li  2 e  0.00000  0.50000  0.50000  1.00000 0,0,0
H1  H   1 d  0.50000  0.50000  0.50000  1.00000 0,0,0
H2  H   1 b  0.00000  0.00000  0.50000  1.00000 0,0,0
F1  F   2 g  0.00000  0.00000  0.31312  1.00000 0,0,Dz
F2  F   4 i  0.00000  0.50000  0.59393  1.00000 0,0,Dz

# end of cif

$KMgF_3$ | LiH | $KMgF_3$ – 25%

# CIF file created by FINDSYM, version 7.1.3

data_findsym-output
_audit_creation_method FINDSYM

_cell_length_a      3.9972000000
_cell_length_b      3.9972000000
_cell_length_c      22.2224000000
_cell_angle_alpha  90.0000000000
_cell_angle_beta   90.0000000000
_cell_angle_gamma  90.0000000000
_cell_volume       355.0607924636

_symmetry_space_group_name_H-M "P 4/m 2/m 2/m"
_symmetry_Int_Tables_number 123
_space_group.reference_setting '123:-P 4 2'
_space_group.transform_Pp_abc a,b,c;0,0,0

loop_
_space_group_symop_id
_space_group_symop_operation_xyz
1 x,y,z
2 x,-y,-z
3 -x,y,-z
4 -x,-y,z
5 -y,-x,-z
6 -y,x,z
7 y,-x,z
8 y,x,-z
9 -x,-y,-z
10 -x,y,z
11 x,-y,z
12 x,y,-z
13 y,x,z
14 y,-x,-z

15 -y,x,-z  
16 -y,-x,z

```
loop_
_atom_site_label
_atom_site_type_symbol
_atom_site_symmetry_multiplicity
_atom_site_Wyckoff_symbol
_atom_site_fract_x
_atom_site_fract_y
_atom_site_fract_z
_atom_site_occupancy
_atom_site_fract_symmform
K1  K   2 h  0.50000  0.50000  0.30210  1.00000  0,0,Dz
Mg1 Mg  2 g  0.00000  0.00000  0.39832  1.00000  0,0,Dz
Li1 Li  2 e  0.00000  0.50000  0.50000  1.00000  0,0,0
H1  H   1 d  0.50000  0.50000  0.50000  1.00000  0,0,0
H2  H   1 b  0.00000  0.00000  0.50000  1.00000  0,0,0
F1  F   2 g  0.00000  0.00000  0.31168  1.00000  0,0,Dz
F2  F   4 i  0.00000  0.50000  0.59016  1.00000  0,0,Dz
```

# end of cif

| LiBaF$_3$ | LiH | LiBaF$_3$ – 0% |

# CIF file created by FINDSYM, version 7.1.3

```
data_findsym-output
_audit_creation_method FINDSYM

_cell_length_a     3.9974000000
_cell_length_b     3.9974000000
_cell_length_c     30.0206000000
_cell_angle_alpha  90.0000000000
_cell_angle_beta   90.0000000000
_cell_angle_gamma  90.0000000000
_cell_volume       479.7053744593

_symmetry_space_group_name_H-M "P 4/m 2/m 2/m"
_symmetry_Int_Tables_number 123
_space_group.reference_setting '123:-P 4 2'
_space_group.transform_Pp_abc a,b,c;0,0,0

loop_
_space_group_symop_id
_space_group_symop_operation_xyz
1 x,y,z
2 x,-y,-z
3 -x,y,-z
4 -x,-y,z
5 -y,-x,-z
6 -y,x,z
```

7 y,-x,z
8 y,x,-z
9 -x,-y,-z
10 -x,y,z
11 x,-y,z
12 x,y,-z
13 y,x,z
14 y,-x,-z
15 -y,x,-z
16 -y,-x,z

loop_
_atom_site_label
_atom_site_type_symbol
_atom_site_symmetry_multiplicity
_atom_site_Wyckoff_symbol
_atom_site_fract_x
_atom_site_fract_y
_atom_site_fract_z
_atom_site_occupancy
_atom_site_fract_symmform
Ba1 Ba  2 g  0.00000  0.00000  0.37321  1.00000 0,0,Dz
Li1 Li  2 h  0.50000  0.50000  0.43684  1.00000 0,0,Dz
Li2 Li  2 e  0.00000  0.50000  0.50000  1.00000 0,0,0
F1  F   2 h  0.50000  0.50000  0.35039  1.00000 0,0,Dz
F2  F   4 i  0.00000  0.50000  0.42670  1.00000 0,0,Dz
H1  H   1 d  0.50000  0.50000  0.50000  1.00000 0,0,0
H2  H   1 b  0.00000  0.00000  0.50000  1.00000 0,0,0

# end of cif

LiBaF$_3$ | LiH | LiBaF$_3$ – 31%

# CIF file created by FINDSYM, version 7.1.3

data_findsym-output
_audit_creation_method FINDSYM

_cell_length_a     3.9974000000
_cell_length_b     3.9974000000
_cell_length_c     30.0206000000
_cell_angle_alpha  90.0000000000
_cell_angle_beta   90.0000000000
_cell_angle_gamma  90.0000000000
_cell_volume       479.7053744593

_symmetry_space_group_name_H-M "P 4/m 2/m 2/m"
_symmetry_Int_Tables_number 123
_space_group.reference_setting '123:-P 4 2'
_space_group.transform_Pp_abc a,b,c;0,0,0

loop_

_space_group_symop_id
_space_group_symop_operation_xyz
1 x,y,z
2 x,-y,-z
3 -x,y,-z
4 -x,-y,z
5 -y,-x,-z
6 -y,x,z
7 y,-x,z
8 y,x,-z
9 -x,-y,-z
10 -x,y,z
11 x,-y,z
12 x,y,-z
13 y,x,z
14 y,-x,-z
15 -y,x,-z
16 -y,-x,z

loop_
_atom_site_label
_atom_site_type_symbol
_atom_site_symmetry_multiplicity
_atom_site_Wyckoff_symbol
_atom_site_fract_x
_atom_site_fract_y
_atom_site_fract_z
_atom_site_occupancy
_atom_site_fract_symmform
Ba1 Ba  2 g  0.00000  0.00000  0.37281  1.00000 0,0,Dz
Li1 Li  2 h  0.50000  0.50000  0.43063  1.00000 0,0,Dz
Li2 Li  2 e  0.00000  0.50000  0.50000  1.00000 0,0,0
F1  F   2 h  0.50000  0.50000  0.36249  1.00000 0,0,Dz
F2  F   4 i  0.00000  0.50000  0.43141  1.00000 0,0,Dz
H1  H   1 d  0.50000  0.50000  0.50000  1.00000 0,0,0
H2  H   1 b  0.00000  0.00000  0.50000  1.00000 0,0,0

# end of cif

$RbMgF_3$ | $MgH_2$ | $RbMgF_3$ – 0%

# CIF file created by FINDSYM, version 7.1.3

data_findsym-output
_audit_creation_method FINDSYM

_cell_length_a    3.9993000000
_cell_length_b    3.9993000000
_cell_length_c   22.2224000000
_cell_angle_alpha 90.0000000000
_cell_angle_beta  90.0000000000
_cell_angle_gamma 90.0000000000

_cell_volume       355.4339654490

_symmetry_space_group_name_H-M "P 4/m 2/m 2/m"
_symmetry_Int_Tables_number 123
_space_group.reference_setting '123:-P 4 2'
_space_group.transform_Pp_abc a,b,c;0,0,0

loop_
_space_group_symop_id
_space_group_symop_operation_xyz
1 x,y,z
2 x,-y,-z
3 -x,y,-z
4 -x,-y,z
5 -y,-x,-z
6 -y,x,z
7 y,-x,z
8 y,x,-z
9 -x,-y,-z
10 -x,y,z
11 x,-y,z
12 x,y,-z
13 y,x,z
14 y,-x,-z
15 -y,x,-z
16 -y,-x,z

loop_
_atom_site_label
_atom_site_type_symbol
_atom_site_symmetry_multiplicity
_atom_site_Wyckoff_symbol
_atom_site_fract_x
_atom_site_fract_y
_atom_site_fract_z
_atom_site_occupancy
_atom_site_fract_symmform
Rb1 Rb   2 h  0.50000  0.50000  0.40315  1.00000 0,0,Dz
Mg1 Mg   1 b  0.00000  0.00000  0.50000  1.00000 0,0,0
Mg2 Mg   2 g  0.00000  0.00000  0.31387  1.00000 0,0,Dz
H1  H    2 e  0.00000  0.50000  0.50000  1.00000 0,0,0
F1  F    2 g  0.00000  0.00000  0.40537  1.00000 0,0,Dz
F2  F    4 i  0.00000  0.50000  0.69258  1.00000 0,0,Dz

# end of cif

$RbMgF_3 \mid MgH_2 \mid RbMgF_3 - 18\%$

# CIF file created by FINDSYM, version 7.1.3

data_findsym-output
_audit_creation_method FINDSYM

```
_cell_length_a     3.9986000000
_cell_length_b     3.9986000000
_cell_length_c     22.2224000000
_cell_angle_alpha  90.0000000000
_cell_angle_beta   90.0000000000
_cell_angle_gamma  90.0000000000
_cell_volume       355.3095526759

_symmetry_space_group_name_H-M "P 4/m 2/m 2/m"
_symmetry_Int_Tables_number 123
_space_group.reference_setting '123:-P 4 2'
_space_group.transform_Pp_abc a,b,c;0,0,0

loop_
_space_group_symop_id
_space_group_symop_operation_xyz
1 x,y,z
2 x,-y,-z
3 -x,y,-z
4 -x,-y,z
5 -y,-x,-z
6 -y,x,z
7 y,-x,z
8 y,x,-z
9 -x,-y,-z
10 -x,y,z
11 x,-y,z
12 x,y,-z
13 y,x,z
14 y,-x,-z
15 -y,x,-z
16 -y,-x,z

loop_
_atom_site_label
_atom_site_type_symbol
_atom_site_symmetry_multiplicity
_atom_site_Wyckoff_symbol
_atom_site_fract_x
_atom_site_fract_y
_atom_site_fract_z
_atom_site_occupancy
_atom_site_fract_symmform
Rb1 Rb   2 h  0.50000  0.50000  0.39701  1.00000 0,0,Dz
Mg1 Mg   1 b  0.00000  0.00000  0.50000  1.00000 0,0,0
Mg2 Mg   2 g  0.00000  0.00000  0.30916  1.00000 0,0,Dz
H1  H    2 e  0.00000  0.50000  0.50000  1.00000 0,0,0
F1  F    2 g  0.00000  0.00000  0.40763  1.00000 0,0,Dz
F2  F    4 i  0.00000  0.50000  0.69217  1.00000 0,0,Dz
```

# end of cif

LiBaF$_3$ | LiBaF$_3$ | LiH | LiBaF$_3$ | LiBaF$_3$ – 0%

# CIF file created by FINDSYM, version 7.1.3

data_findsym-output
_audit_creation_method FINDSYM

_cell_length_a     4.0026000000
_cell_length_b     4.0026000000
_cell_length_c     40.5574000000
_cell_angle_alpha  90.0000000000
_cell_angle_beta   90.0000000000
_cell_angle_gamma  90.0000000000
_cell_volume       649.7622680880

_symmetry_space_group_name_H-M "P 4/m 2/m 2/m"
_symmetry_Int_Tables_number 123
_space_group.reference_setting '123:-P 4 2'
_space_group.transform_Pp_abc a,b,c;0,0,0

loop_
_space_group_symop_id
_space_group_symop_operation_xyz
1 x,y,z
2 x,-y,-z
3 -x,y,-z
4 -x,-y,z
5 -y,-x,-z
6 -y,x,z
7 y,-x,z
8 y,x,-z
9 -x,-y,-z
10 -x,y,z
11 x,-y,z
12 x,y,-z
13 y,x,z
14 y,-x,-z
15 -y,x,-z
16 -y,-x,z

loop_
_atom_site_label
_atom_site_type_symbol
_atom_site_symmetry_multiplicity
_atom_site_Wyckoff_symbol
_atom_site_fract_x
_atom_site_fract_y
_atom_site_fract_z
_atom_site_occupancy

_atom_site_fract_symmform
Ba1 Ba  2 g  0.00000  0.00000  0.70507  1.00000 0,0,Dz
Ba2 Ba  2 g  0.00000  0.00000  0.59385  1.00000 0,0,Dz
Li1 Li  2 e  0.00000  0.50000  0.50000  1.00000 0,0,0
Li2 Li  2 h  0.50000  0.50000  0.65378  1.00000 0,0,Dz
Li3 Li  2 h  0.50000  0.50000  0.54667  1.00000 0,0,Dz
F1  F   4 i  0.00000  0.50000  0.66444  1.00000 0,0,Dz
F2  F   2 h  0.50000  0.50000  0.60843  1.00000 0,0,Dz
F3  F   4 i  0.00000  0.50000  0.55423  1.00000 0,0,Dz
F4  F   2 h  0.50000  0.50000  0.27728  1.00000 0,0,Dz
H1  H   1 d  0.50000  0.50000  0.50000  1.00000 0,0,0
H2  H   1 b  0.00000  0.00000  0.50000  1.00000 0,0,0

# end of cif

$LiBaF_3$ | $LiBaF_3$ | LiH | $LiBaF_3$ | $LiBaF_3$ – 31%

# CIF file created by FINDSYM, version 7.1.3

data_findsym-output
_audit_creation_method FINDSYM

_cell_length_a     4.0026000000
_cell_length_b     4.0026000000
_cell_length_c     40.5574000000
_cell_angle_alpha  90.0000000000
_cell_angle_beta   90.0000000000
_cell_angle_gamma  90.0000000000
_cell_volume       649.7622680880

_symmetry_space_group_name_H-M "P 4/m 2/m 2/m"
_symmetry_Int_Tables_number 123
_space_group.reference_setting '123:-P 4 2'
_space_group.transform_Pp_abc a,b,c;0,0,0

loop_
_space_group_symop_id
_space_group_symop_operation_xyz
1 x,y,z
2 x,-y,-z
3 -x,y,-z
4 -x,-y,z
5 -y,-x,-z
6 -y,x,z
7 y,-x,z
8 y,x,-z
9 -x,-y,-z
10 -x,y,z
11 x,-y,z
12 x,y,-z
13 y,x,z
14 y,-x,-z

```
15 -y,x,-z
16 -y,-x,z

loop_
_atom_site_label
_atom_site_type_symbol
_atom_site_symmetry_multiplicity
_atom_site_Wyckoff_symbol
_atom_site_fract_x
_atom_site_fract_y
_atom_site_fract_z
_atom_site_occupancy
_atom_site_fract_symmform
Ba1 Ba   2 g  0.00000  0.00000  0.69780  1.00000 0,0,Dz
Ba2 Ba   2 g  0.00000  0.00000  0.59385  1.00000 0,0,Dz
Li1 Li   2 e  0.00000  0.50000  0.50000  1.00000 0,0,0
Li2 Li   2 h  0.50000  0.50000  0.64874  1.00000 0,0,Dz
Li3 Li   2 h  0.50000  0.50000  0.55037  1.00000 0,0,Dz
F1  F    4 i  0.00000  0.50000  0.65379  1.00000 0,0,Dz
F2  F    2 h  0.50000  0.50000  0.60088  1.00000 0,0,Dz
F3  F    4 i  0.00000  0.50000  0.55013  1.00000 0,0,Dz
F4  F    2 h  0.50000  0.50000  0.29284  1.00000 0,0,Dz
H1  H    1 d  0.50000  0.50000  0.50000  1.00000 0,0,0
H2  H    1 b  0.00000  0.00000  0.50000  1.00000 0,0,0

# end of cif
```

$KMgF_3$ | $KMgF_3$ | LiH | $KMgF_3$ | $KMgF_3$ – 0%

```
# CIF file created by FINDSYM, version 7.1.3

data_findsym-output
_audit_creation_method FINDSYM

_cell_length_a    4.0183000000
_cell_length_b    4.0183000000
_cell_length_c    32.0240000000
_cell_angle_alpha 90.0000000000
_cell_angle_beta  90.0000000000
_cell_angle_gamma 90.0000000000
_cell_volume      517.0830381174

_symmetry_space_group_name_H-M "P 4/m 2/m 2/m"
_symmetry_Int_Tables_number 123
_space_group.reference_setting '123:-P 4 2'
_space_group.transform_Pp_abc a,b,c;0,0,0

loop_
_space_group_symop_id
_space_group_symop_operation_xyz
1 x,y,z
2 x,-y,-z
```

3 -x,y,-z
4 -x,-y,z
5 -y,-x,-z
6 -y,x,z
7 y,-x,z
8 y,x,-z
9 -x,-y,-z
10 -x,y,z
11 x,-y,z
12 x,y,-z
13 y,x,z
14 y,-x,-z
15 -y,x,-z
16 -y,-x,z

loop_
_atom_site_label
_atom_site_type_symbol
_atom_site_symmetry_multiplicity
_atom_site_Wyckoff_symbol
_atom_site_fract_x
_atom_site_fract_y
_atom_site_fract_z
_atom_site_occupancy
_atom_site_fract_symmform
K1  K   2 g  0.00000  0.00000  0.24620  1.00000 0,0,Dz
K2  K   2 g  0.00000  0.00000  0.37258  1.00000 0,0,Dz
Mg1 Mg  2 h  0.50000  0.50000  0.30672  1.00000 0,0,Dz
Mg2 Mg  2 h  0.50000  0.50000  0.43285  1.00000 0,0,Dz
F1  F   2 h  0.50000  0.50000  0.24405  1.00000 0,0,Dz
F2  F   4 i  0.00000  0.50000  0.30762  1.00000 0,0,Dz
F3  F   2 h  0.50000  0.50000  0.37061  1.00000 0,0,Dz
F4  F   4 i  0.00000  0.50000  0.43494  1.00000 0,0,Dz
H1  H   1 b  0.00000  0.00000  0.50000  1.00000 0,0,0
H2  H   1 d  0.50000  0.50000  0.50000  1.00000 0,0,0
Li1 Li  2 e  0.00000  0.50000  0.50000  1.00000 0,0,0

# end of cif

$KMgF_3$ | $KMgF_3$ | LiH | $KMgF_3$ | $KMgF_3$ – 20%

# CIF file created by FINDSYM, version 7.1.3

data_findsym-output
_audit_creation_method FINDSYM

_cell_length_a    4.0183000000
_cell_length_b    4.0183000000
_cell_length_c    32.0240000000
_cell_angle_alpha 90.0000000000
_cell_angle_beta  90.0000000000
_cell_angle_gamma 90.0000000000

```
_cell_volume       517.0830381174

_symmetry_space_group_name_H-M "P 4/m 2/m 2/m"
_symmetry_Int_Tables_number 123
_space_group.reference_setting '123:-P 4 2'
_space_group.transform_Pp_abc a,b,c;0,0,0

loop_
_space_group_symop_id
_space_group_symop_operation_xyz
1 x,y,z
2 x,-y,-z
3 -x,y,-z
4 -x,-y,z
5 -y,-x,-z
6 -y,x,z
7 y,-x,z
8 y,x,-z
9 -x,-y,-z
10 -x,y,z
11 x,-y,z
12 x,y,-z
13 y,x,z
14 y,-x,-z
15 -y,x,-z
16 -y,-x,z

loop_
_atom_site_label
_atom_site_type_symbol
_atom_site_symmetry_multiplicity
_atom_site_Wyckoff_symbol
_atom_site_fract_x
_atom_site_fract_y
_atom_site_fract_z
_atom_site_occupancy
_atom_site_fract_symmform
K1  K   2 g  0.00000  0.00000  0.23772  1.00000 0,0,Dz
K2  K   2 g  0.00000  0.00000  0.36576  1.00000 0,0,Dz
Mg1 Mg  2 h  0.50000  0.50000  0.30273  1.00000 0,0,Dz
Mg2 Mg  2 h  0.50000  0.50000  0.43028  1.00000 0,0,Dz
F1  F   2 h  0.50000  0.50000  0.24151  1.00000 0,0,Dz
F2  F   4 i  0.00000  0.50000  0.30622  1.00000 0,0,Dz
F3  F   2 h  0.50000  0.50000  0.36973  1.00000 0,0,Dz
F4  F   4 i  0.00000  0.50000  0.43731  1.00000 0,0,Dz
H1  H   1 b  0.00000  0.00000  0.50000  1.00000 0,0,0
H2  H   1 d  0.50000  0.50000  0.50000  1.00000 0,0,0
Li1 Li  2 e  0.00000  0.50000  0.50000  1.00000 0,0,0

# end of cif
```

| RbMgF$_3$ \| LiH \| RbMgF$_3$  - 0% |
|---|
| # CIF file created by FINDSYM, version 7.1.3<br><br>data_findsym-output<br>_audit_creation_method FINDSYM<br><br>_cell_length_a    4.0514000000<br>_cell_length_b    4.0514000000<br>_cell_length_c    22.2224000000<br>_cell_angle_alpha 90.0000000000<br>_cell_angle_beta  90.0000000000<br>_cell_angle_gamma 90.0000000000<br>_cell_volume      364.7549615719<br><br>_symmetry_space_group_name_H-M "P 4/m 2/m 2/m"<br>_symmetry_Int_Tables_number 123<br>_space_group.reference_setting '123:-P 4 2'<br>_space_group.transform_Pp_abc a,b,c;0,0,0<br><br>loop_<br>_space_group_symop_id<br>_space_group_symop_operation_xyz<br>1 x,y,z<br>2 x,-y,-z<br>3 -x,y,-z<br>4 -x,-y,z<br>5 -y,-x,-z<br>6 -y,x,z<br>7 y,-x,z<br>8 y,x,-z<br>9 -x,-y,-z<br>10 -x,y,z<br>11 x,-y,z<br>12 x,y,-z<br>13 y,x,z<br>14 y,-x,-z<br>15 -y,x,-z<br>16 -y,-x,z<br><br>loop_<br>_atom_site_label<br>_atom_site_type_symbol<br>_atom_site_symmetry_multiplicity<br>_atom_site_Wyckoff_symbol<br>_atom_site_fract_x<br>_atom_site_fract_y<br>_atom_site_fract_z<br>_atom_site_occupancy<br>_atom_site_fract_symmform<br>Rb1 Rb  2 h  0.50000  0.50000  0.31335  1.00000 0,0,Dz |
| Mg1 Mg   2 g  0.00000  0.00000  0.40333  1.00000 0,0,Dz<br>Li1 Li   2 e  0.00000  0.50000  0.50000  1.00000 0,0,0<br>H1  H    1 d  0.50000  0.50000  0.50000  1.00000 0,0,0<br>H2  H    1 b  0.00000  0.00000  0.50000  1.00000 0,0,0<br>F1  F    2 g  0.00000  0.00000  0.31379  1.00000 0,0,Dz<br>F2  F    4 i  0.00000  0.50000  0.59299  1.00000 0,0,Dz<br><br># end of cif |
| $RbMgF_3$ \| LiH \| $RbMgF_3 - 25\%$ |
| # CIF file created by FINDSYM, version 7.1.3<br><br>data_findsym-output<br>_audit_creation_method FINDSYM<br><br>_cell_length_a     3.9986000000<br>_cell_length_b     3.9986000000<br>_cell_length_c     22.2224000000<br>_cell_angle_alpha  90.0000000000<br>_cell_angle_beta   90.0000000000<br>_cell_angle_gamma  90.0000000000<br>_cell_volume       355.3095526759<br><br>_symmetry_space_group_name_H-M "P 4/m 2/m 2/m"<br>_symmetry_Int_Tables_number 123<br>_space_group.reference_setting '123:-P 4 2'<br>_space_group.transform_Pp_abc a,b,c;0,0,0<br><br>loop_<br>_space_group_symop_id<br>_space_group_symop_operation_xyz<br>1 x,y,z<br>2 x,-y,-z<br>3 -x,y,-z<br>4 -x,-y,z<br>5 -y,-x,-z<br>6 -y,x,z<br>7 y,-x,z<br>8 y,x,-z<br>9 -x,-y,-z<br>10 -x,y,z<br>11 x,-y,z<br>12 x,y,-z<br>13 y,x,z<br>14 y,-x,-z<br>15 -y,x,-z<br>16 -y,-x,z<br><br>loop_<br>_atom_site_label<br>_atom_site_type_symbol |

_atom_site_symmetry_multiplicity
_atom_site_Wyckoff_symbol
_atom_site_fract_x
_atom_site_fract_y
_atom_site_fract_z
_atom_site_occupancy
_atom_site_fract_symmform
Rb1 Rb  2 h  0.50000  0.50000  0.39701  1.00000 0,0,Dz
Mg1 Mg  1 b  0.00000  0.00000  0.50000  1.00000 0,0,0
Mg2 Mg  2 g  0.00000  0.00000  0.30916  1.00000 0,0,Dz
H1  H   2 e  0.00000  0.50000  0.50000  1.00000 0,0,0
F1  F   2 g  0.00000  0.00000  0.40763  1.00000 0,0,Dz
F2  F   4 i  0.00000  0.50000  0.69217  1.00000 0,0,Dz

# end of cif

$RbMgF_3$ | $RbMgF_3$ | LiH | $RbMgF_3$ | $RbMgF_3$ - 0%
# CIF file created by FINDSYM, version 7.1.3

data_findsym-output
_audit_creation_method FINDSYM

_cell_length_a     4.0813000000
_cell_length_b     4.0813000000
_cell_length_c     32.0800000000
_cell_angle_alpha  90.0000000000
_cell_angle_beta   90.0000000000
_cell_angle_gamma  90.0000000000
_cell_volume       534.3568708552

_symmetry_space_group_name_H-M "P 4/m 2/m 2/m"
_symmetry_Int_Tables_number 123
_space_group.reference_setting '123:-P 4 2'
_space_group.transform_Pp_abc a,b,c;0,0,0

loop_
_space_group_symop_id
_space_group_symop_operation_xyz
1 x,y,z
2 x,-y,-z
3 -x,y,-z
4 -x,-y,z
5 -y,-x,-z
6 -y,x,z
7 y,-x,z
8 y,x,-z
9 -x,-y,-z
10 -x,y,z
11 x,-y,z
12 x,y,-z
13 y,x,z

14 y,-x,-z
15 -y,x,-z
16 -y,-x,z

loop_
_atom_site_label
_atom_site_type_symbol
_atom_site_symmetry_multiplicity
_atom_site_Wyckoff_symbol
_atom_site_fract_x
_atom_site_fract_y
_atom_site_fract_z
_atom_site_occupancy
_atom_site_fract_symmform
Rb1 Rb   2 g  0.00000  0.00000  0.24014  1.00000  0,0,Dz
Rb2 Rb   2 g  0.00000  0.00000  0.37171  1.00000  0,0,Dz
Mg1 Mg   2 h  0.50000  0.50000  0.30513  1.00000  0,0,Dz
Mg2 Mg   2 h  0.50000  0.50000  0.43329  1.00000  0,0,Dz
F1  F    2 h  0.50000  0.50000  0.24229  1.00000  0,0,Dz
F2  F    4 i  0.00000  0.50000  0.30544  1.00000  0,0,Dz
F3  F    2 h  0.50000  0.50000  0.37008  1.00000  0,0,Dz
F4  F    4 i  0.00000  0.50000  0.43587  1.00000  0,0,Dz
H1  H    1 b  0.00000  0.00000  0.50000  1.00000  0,0,0
H2  H    1 d  0.50000  0.50000  0.50000  1.00000  0,0,0
Li1 Li   2 e  0.00000  0.50000  0.50000  1.00000  0,0,0

# end of cif

$RbMgF_3$ | $RbMgF_3$ | LiH | $RbMgF_3$ | $RbMgF_3$ - 27%

# CIF file created by FINDSYM, version 7.1.3

data_findsym-output
_audit_creation_method FINDSYM

_cell_length_a     4.0813000000
_cell_length_b     4.0813000000
_cell_length_c     32.0800000000
_cell_angle_alpha  90.0000000000
_cell_angle_beta   90.0000000000
_cell_angle_gamma  90.0000000000
_cell_volume       534.3568708552

_symmetry_space_group_name_H-M "P 4/m 2/m 2/m"
_symmetry_Int_Tables_number 123
_space_group.reference_setting '123:-P 4 2'
_space_group.transform_Pp_abc a,b,c;0,0,0

loop_
_space_group_symop_id
_space_group_symop_operation_xyz
1 x,y,z

2 x,-y,-z
3 -x,y,-z
4 -x,-y,z
5 -y,-x,-z
6 -y,x,z
7 y,-x,z
8 y,x,-z
9 -x,-y,-z
10 -x,y,z
11 x,-y,z
12 x,y,-z
13 y,x,z
14 y,-x,-z
15 -y,x,-z
16 -y,-x,z

loop_
_atom_site_label
_atom_site_type_symbol
_atom_site_symmetry_multiplicity
_atom_site_Wyckoff_symbol
_atom_site_fract_x
_atom_site_fract_y
_atom_site_fract_z
_atom_site_occupancy
_atom_site_fract_symmform
Rb1 Rb   2 g  0.00000  0.00000  0.22845  1.00000 0,0,Dz
Rb2 Rb   2 g  0.00000  0.00000  0.36445  1.00000 0,0,Dz
Mg1 Mg   2 h  0.50000  0.50000  0.29874  1.00000 0,0,Dz
Mg2 Mg   2 h  0.50000  0.50000  0.43010  1.00000 0,0,Dz
F1  F    2 h  0.50000  0.50000  0.23794  1.00000 0,0,Dz
F2  F    4 i  0.00000  0.50000  0.30286  1.00000 0,0,Dz
F3  F    2 h  0.50000  0.50000  0.36977  1.00000 0,0,Dz
F4  F    4 i  0.00000  0.50000  0.43897  1.00000 0,0,Dz
H1  H    1 b  0.00000  0.00000  0.50000  1.00000 0,0,0
H2  H    1 d  0.50000  0.50000  0.50000  1.00000 0,0,0
Li1 Li   2 e  0.00000  0.50000  0.50000  1.00000 0,0,0

# end of cif

$KZnF_3$ | $KZnF_3$ | LiH | $KZnF_3$ | $KZnF_3$ – 0%

# CIF file created by FINDSYM, version 7.1.3

data_findsym-output
_audit_creation_method FINDSYM

_cell_length_a     4.1314000000
_cell_length_b     4.1314000000
_cell_length_c     32.2473000000
_cell_angle_alpha  90.0000000000
_cell_angle_beta   90.0000000000

```
_cell_angle_gamma  90.0000000000
_cell_volume       550.4119423519

_symmetry_space_group_name_H-M "P 4/m 2/m 2/m"
_symmetry_Int_Tables_number 123
_space_group.reference_setting '123:-P 4 2'
_space_group.transform_Pp_abc a,b,c;0,0,0

loop_
_space_group_symop_id
_space_group_symop_operation_xyz
1 x,y,z
2 x,-y,-z
3 -x,y,-z
4 -x,-y,z
5 -y,-x,-z
6 -y,x,z
7 y,-x,z
8 y,x,-z
9 -x,-y,-z
10 -x,y,z
11 x,-y,z
12 x,y,-z
13 y,x,z
14 y,-x,-z
15 -y,x,-z
16 -y,-x,z

loop_
_atom_site_label
_atom_site_type_symbol
_atom_site_symmetry_multiplicity
_atom_site_Wyckoff_symbol
_atom_site_fract_x
_atom_site_fract_y
_atom_site_fract_z
_atom_site_occupancy
_atom_site_fract_symmform
K1  K   2 g  0.00000  0.00000  0.25503  1.00000 0,0,Dz
K2  K   2 g  0.00000  0.00000  0.37825  1.00000 0,0,Dz
Zn1 Zn  2 h  0.50000  0.50000  0.31223  1.00000 0,0,Dz
Zn2 Zn  2 h  0.50000  0.50000  0.43977  1.00000 0,0,Dz
F1  F   2 h  0.50000  0.50000  0.24988  1.00000 0,0,Dz
F2  F   4 i  0.00000  0.50000  0.31372  1.00000 0,0,Dz
F3  F   2 h  0.50000  0.50000  0.37671  1.00000 0,0,Dz
F4  F   4 i  0.00000  0.50000  0.43874  1.00000 0,0,Dz
H1  H   1 b  0.00000  0.00000  0.50000  1.00000 0,0,0
H2  H   1 d  0.50000  0.50000  0.50000  1.00000 0,0,0
Li1 Li  2 e  0.00000  0.50000  0.50000  1.00000 0,0,0
```

# end of cif

KZnF$_3$ | KZnF$_3$ | LiH | KZnF$_3$ | KZnF$_3$ – 5%

# CIF file created by FINDSYM, version 7.1.3

data_findsym-output
_audit_creation_method FINDSYM

_cell_length_a     4.1314000000
_cell_length_b     4.1314000000
_cell_length_c     32.2473000000
_cell_angle_alpha  90.0000000000
_cell_angle_beta   90.0000000000
_cell_angle_gamma  90.0000000000
_cell_volume       550.4119423519

_symmetry_space_group_name_H-M "P 4/m 2/m 2/m"
_symmetry_Int_Tables_number 123
_space_group.reference_setting '123:-P 4 2'
_space_group.transform_Pp_abc a,b,c;0,0,0

loop_
_space_group_symop_id
_space_group_symop_operation_xyz
1 x,y,z
2 x,-y,-z
3 -x,y,-z
4 -x,-y,z
5 -y,-x,-z
6 -y,x,z
7 y,-x,z
8 y,x,-z
9 -x,-y,-z
10 -x,y,z
11 x,-y,z
12 x,y,-z
13 y,x,z
14 y,-x,-z
15 -y,x,-z
16 -y,-x,z

loop_
_atom_site_label
_atom_site_type_symbol
_atom_site_symmetry_multiplicity
_atom_site_Wyckoff_symbol
_atom_site_fract_x
_atom_site_fract_y
_atom_site_fract_z
_atom_site_occupancy
_atom_site_fract_symmform

| | |
|---|---|
| K1  K   2 g  0.00000  0.00000  0.25347  1.00000 0,0,Dz
K2  K   2 g  0.00000  0.00000  0.37690  1.00000 0,0,Dz
Zn1 Zn  2 h  0.50000  0.50000  0.31158  1.00000 0,0,Dz
Zn2 Zn  2 h  0.50000  0.50000  0.43913  1.00000 0,0,Dz
F1  F   2 h  0.50000  0.50000  0.24963  1.00000 0,0,Dz
F2  F   4 i  0.00000  0.50000  0.31372  1.00000 0,0,Dz
F3  F   2 h  0.50000  0.50000  0.37669  1.00000 0,0,Dz
F4  F   4 i  0.00000  0.50000  0.43928  1.00000 0,0,Dz
H1  H   1 b  0.00000  0.00000  0.50000  1.00000 0,0,0
H2  H   1 d  0.50000  0.50000  0.50000  1.00000 0,0,0
Li1 Li  2 e  0.00000  0.50000  0.50000  1.00000 0,0,0

# end of cif | |
| $KZnF_3$ \| $LiH$ \| $KZnF_3$ – 0% | |
| # CIF file created by FINDSYM, version 7.1.3

data_findsym-output
_audit_creation_method FINDSYM

 _cell_length_a    5.9184000000
 _cell_length_b    5.9184000000
 _cell_length_c    23.7638000000
 _cell_angle_alpha 90.0000000000
 _cell_angle_beta  90.0000000000
 _cell_angle_gamma 90.0000000000
 _cell_volume      832.3855197281

_symmetry_space_group_name_H-M "P 4/n $2_1$/m 2/m (origin choice 2)"
_symmetry_Int_Tables_number 129
_space_group.reference_setting '129:-P 4a 2a'
_space_group.transform_Pp_abc a,b,c;0,0,0

loop_
_space_group_symop_id
_space_group_symop_operation_xyz
1 x,y,z
2 x+1/2,-y,-z
3 -x,y+1/2,-z
4 -x+1/2,-y+1/2,z
5 -y,-x,-z
6 -y+1/2,x,z
7 y,-x+1/2,z
8 y+1/2,x+1/2,-z
9 -x,-y,-z
10 -x+1/2,y,z
11 x,-y+1/2,z
12 x+1/2,y+1/2,-z
13 y,x,z
14 y+1/2,-x,-z
15 -y,x+1/2,-z | |

16 -y+1/2,-x+1/2,z

loop_
_atom_site_label
_atom_site_type_symbol
_atom_site_symmetry_multiplicity
_atom_site_Wyckoff_symbol
_atom_site_fract_x
_atom_site_fract_y
_atom_site_fract_z
_atom_site_occupancy
_atom_site_fract_symmform
K1   K    4 f  0.75000  0.25000  0.65891  1.00000  0,0,Dz
Zn1  Zn   2 c  0.25000  0.25000  0.59467  1.00000  0,0,Dz
Zn2  Zn   2 c  0.25000  0.25000  0.41903  1.00000  0,0,Dz
F1   F    2 c  0.25000  0.25000  0.67745  1.00000  0,0,Dz
F2   F    2 c  0.25000  0.25000  0.33736  1.00000  0,0,Dz
F3   F    8 j  0.01112  0.01112  0.58439  1.00000  Dx,Dx,Dz
H1   H    2 b  0.75000  0.25000  0.50000  1.00000  0,0,0
H2   H    2 c  0.25000  0.25000  0.48891  1.00000  0,0,Dz
Li1  Li   4 e  0.00000  0.00000  0.50000  1.00000  0,0,0

# end of cif

$KZnF_3$ | LiH | $KZnF_3$ – 5%

# CIF file created by FINDSYM, version 7.1.3

data_findsym-output
_audit_creation_method FINDSYM

_cell_length_a     5.9184000000
_cell_length_b     5.9184000000
_cell_length_c     23.7638000000
_cell_angle_alpha  90.0000000000
_cell_angle_beta   90.0000000000
_cell_angle_gamma  90.0000000000
_cell_volume       832.3855197281

_symmetry_space_group_name_H-M "P 4/n 2$_1$/m 2/m (origin choice 2)"
_symmetry_Int_Tables_number 129
_space_group.reference_setting '129:-P 4a 2a'
_space_group.transform_Pp_abc a,b,c;0,0,0

loop_
_space_group_symop_id
_space_group_symop_operation_xyz
1 x,y,z
2 x+1/2,-y,-z
3 -x,y+1/2,-z
4 -x+1/2,-y+1/2,z
5 -y,-x,-z

6 -y+1/2,x,z
7 y,-x+1/2,z
8 y+1/2,x+1/2,-z
9 -x,-y,-z
10 -x+1/2,y,z
11 x,-y+1/2,z
12 x+1/2,y+1/2,-z
13 y,x,z
14 y+1/2,-x,-z
15 -y,x+1/2,-z
16 -y+1/2,-x+1/2,z

loop_
_atom_site_label
_atom_site_type_symbol
_atom_site_symmetry_multiplicity
_atom_site_Wyckoff_symbol
_atom_site_fract_x
_atom_site_fract_y
_atom_site_fract_z
_atom_site_occupancy
_atom_site_fract_symmform
K1  K   4 f  0.75000  0.25000  0.66086  1.00000  0,0,Dz
Zn1 Zn  2 c  0.25000  0.25000  0.59465  1.00000  0,0,Dz
Zn2 Zn  2 c  0.25000  0.25000  0.41886  1.00000  0,0,Dz
F1  F   2 c  0.25000  0.25000  0.67686  1.00000  0,0,Dz
F2  F   2 c  0.25000  0.25000  0.33761  1.00000  0,0,Dz
F3  F   8 j  0.01098  0.01098  0.58351  1.00000  Dx,Dx,Dz
H1  H   2 b  0.75000  0.25000  0.50000  1.00000  0,0,0
H2  H   2 c  0.25000  0.25000  0.48916  1.00000  0,0,Dz
Li1 Li  4 e  0.00000  0.00000  0.50000  1.00000  0,0,0

# end of cif

CaO | CaO | NaH | CaO | CaO – 0%

# CIF file created by FINDSYM, version 7.1.3

data_findsym-output
_audit_creation_method FINDSYM

_cell_length_a     3.3512000000
_cell_length_b     3.3512000000
_cell_length_c     29.6096000000
_cell_angle_alpha  90.0000000000
_cell_angle_beta   90.0000000000
_cell_angle_gamma  90.0000000000
_cell_volume       332.5318398218

_symmetry_space_group_name_H-M "P 4/m 2/m 2/m"
_symmetry_Int_Tables_number 123
_space_group.reference_setting '123:-P 4 2'

_space_group.transform_Pp_abc a,b,c;0,0,0

loop_
_space_group_symop_id
_space_group_symop_operation_xyz
1 x,y,z
2 x,-y,-z
3 -x,y,-z
4 -x,-y,z
5 -y,-x,-z
6 -y,x,z
7 y,-x,z
8 y,x,-z
9 -x,-y,-z
10 -x,y,z
11 x,-y,z
12 x,y,-z
13 y,x,z
14 y,-x,-z
15 -y,x,-z
16 -y,-x,z

loop_
_atom_site_label
_atom_site_type_symbol
_atom_site_symmetry_multiplicity
_atom_site_Wyckoff_symbol
_atom_site_fract_x
_atom_site_fract_y
_atom_site_fract_z
_atom_site_occupancy
_atom_site_fract_symmform
Ca1  Ca   2 g   0.00000   0.00000   0.33581   1.00000 0,0,Dz
Ca2  Ca   2 h   0.50000   0.50000   0.41641   1.00000 0,0,Dz
O1   O    2 h   0.50000   0.50000   0.33667   1.00000 0,0,Dz
O2   O    2 g   0.00000   0.00000   0.41592   1.00000 0,0,Dz
Na1  Na   1 b   0.00000   0.00000   0.50000   1.00000 0,0,0
H1   H    1 d   0.50000   0.50000   0.50000   1.00000 0,0,0

# end of cif

| CaO | CaO | NaH | CaO | CaO – 35% |

# CIF file created by FINDSYM, version 7.1.3

data_findsym-output
_audit_creation_method FINDSYM

_cell_length_a     3.3512000000
_cell_length_b     3.3512000000
_cell_length_c    29.6096000000
_cell_angle_alpha 90.0000000000

_cell_angle_beta   90.0000000000
_cell_angle_gamma  90.0000000000
_cell_volume       332.5318398218

_symmetry_space_group_name_H-M "P 4/m 2/m 2/m"
_symmetry_Int_Tables_number 123
_space_group.reference_setting '123:-P 4 2'
_space_group.transform_Pp_abc a,b,c;0,0,0

loop_
_space_group_symop_id
_space_group_symop_operation_xyz
1 x,y,z
2 x,-y,-z
3 -x,y,-z
4 -x,-y,z
5 -y,-x,-z
6 -y,x,z
7 y,-x,z
8 y,x,-z
9 -x,-y,-z
10 -x,y,z
11 x,-y,z
12 x,y,-z
13 y,x,z
14 y,-x,-z
15 -y,x,-z
16 -y,-x,z

loop_
_atom_site_label
_atom_site_type_symbol
_atom_site_symmetry_multiplicity
_atom_site_Wyckoff_symbol
_atom_site_fract_x
_atom_site_fract_y
_atom_site_fract_z
_atom_site_occupancy
_atom_site_fract_symmform
Ca1  Ca   2 g   0.00000  0.00000  0.33167  1.00000 0,0,Dz
Ca2  Ca   2 h   0.50000  0.50000  0.41481  1.00000 0,0,Dz
O1   O    2 h   0.50000  0.50000  0.33589  1.00000 0,0,Dz
O2   O    2 g   0.00000  0.00000  0.41392  1.00000 0,0,Dz
Na1  Na   1 b   0.00000  0.00000  0.50000  1.00000 0,0,0
H1   H    1 d   0.50000  0.50000  0.50000  1.00000 0,0,0

# end of cif

YN | YN | NaH | YN | YN – 0%
# CIF file created by FINDSYM, version 7.1.3

```
data_findsym-output
_audit_creation_method FINDSYM

_cell_length_a    3.4128000000
_cell_length_b    3.4128000000
_cell_length_c    30.0283000000
_cell_angle_alpha 90.0000000000
_cell_angle_beta  90.0000000000
_cell_angle_gamma 90.0000000000
_cell_volume      349.7457310687

_symmetry_space_group_name_H-M "P 4/m 2/m 2/m"
_symmetry_Int_Tables_number 123
_space_group.reference_setting '123:-P 4 2'
_space_group.transform_Pp_abc a,b,c;0,0,0

loop_
_space_group_symop_id
_space_group_symop_operation_xyz
1 x,y,z
2 x,-y,-z
3 -x,y,-z
4 -x,-y,z
5 -y,-x,-z
6 -y,x,z
7 y,-x,z
8 y,x,-z
9 -x,-y,-z
10 -x,y,z
11 x,-y,z
12 x,y,-z
13 y,x,z
14 y,-x,-z
15 -y,x,-z
16 -y,-x,z

loop_
_atom_site_label
_atom_site_type_symbol
_atom_site_symmetry_multiplicity
_atom_site_Wyckoff_symbol
_atom_site_fract_x
_atom_site_fract_y
_atom_site_fract_z
_atom_site_occupancy
_atom_site_fract_symmform
Y1  Y   2 g  0.00000  0.00000  0.33720  1.00000 0,0,Dz
Y2  Y   2 h  0.50000  0.50000  0.41642  1.00000 0,0,Dz
N1  N   2 h  0.50000  0.50000  0.33917  1.00000 0,0,Dz
N2  N   2 g  0.00000  0.00000  0.41475  1.00000 0,0,Dz
```

| |
|---|
| Na1 Na  1 b  0.00000  0.00000  0.50000  1.00000 0,0,0 <br> H1  H   1 d  0.50000  0.50000  0.50000  1.00000 0,0,0 <br> <br> # end of cif |
| YN \| YN \| NaH \| YN \| YN – 72% |
| # CIF file created by FINDSYM, version 7.1.3 <br> <br> data_findsym-output <br> _audit_creation_method FINDSYM <br> <br> _cell_length_a     3.4128000000 <br> _cell_length_b     3.4128000000 <br> _cell_length_c     30.0283000000 <br> _cell_angle_alpha  90.0000000000 <br> _cell_angle_beta   90.0000000000 <br> _cell_angle_gamma  90.0000000000 <br> _cell_volume      349.7457310687 <br> <br> _symmetry_space_group_name_H-M "P 4/m 2/m 2/m" <br> _symmetry_Int_Tables_number 123 <br> _space_group.reference_setting '123:-P 4 2' <br> _space_group.transform_Pp_abc a,b,c;0,0,0 <br> <br> loop_ <br> _space_group_symop_id <br> _space_group_symop_operation_xyz <br> 1 x,y,z <br> 2 x,-y,-z <br> 3 -x,y,-z <br> 4 -x,-y,z <br> 5 -y,-x,-z <br> 6 -y,x,z <br> 7 y,-x,z <br> 8 y,x,-z <br> 9 -x,-y,-z <br> 10 -x,y,z <br> 11 x,-y,z <br> 12 x,y,-z <br> 13 y,x,z <br> 14 y,-x,-z <br> 15 -y,x,-z <br> 16 -y,-x,z <br> <br> loop_ <br> _atom_site_label <br> _atom_site_type_symbol <br> _atom_site_symmetry_multiplicity <br> _atom_site_Wyckoff_symbol <br> _atom_site_fract_x <br> _atom_site_fract_y |

_atom_site_fract_z
_atom_site_occupancy
_atom_site_fract_symmform
Y1  Y   2 g  0.00000  0.00000  0.32586  1.00000 0,0,Dz
Y2  Y   2 h  0.50000  0.50000  0.40913  1.00000 0,0,Dz
N1  N   2 h  0.50000  0.50000  0.33224  1.00000 0,0,Dz
N2  N   2 g  0.00000  0.00000  0.40657  1.00000 0,0,Dz
Na1 Na  1 b  0.00000  0.00000  0.50000  1.00000 0,0,0
H1  H   1 d  0.50000  0.50000  0.50000  1.00000 0,0,0

# end of cif

LiCl | LiCl | NaH | LiCl | LiCl – 0%

# CIF file created by FINDSYM, version 7.1.3

data_findsym-output
_audit_creation_method FINDSYM

_cell_length_a     4.9392000000
_cell_length_b     4.9392000000
_cell_length_c     32.4643000000
_cell_angle_alpha  90.0000000000
_cell_angle_beta   90.0000000000
_cell_angle_gamma  90.0000000000
_cell_volume       791.9892144300

_symmetry_space_group_name_H-M "P 4/n 2$_1$/m 2/m (origin choice 2)"
_symmetry_Int_Tables_number 129
_space_group.reference_setting '129:-P 4a 2a'
_space_group.transform_Pp_abc a,b,c;0,0,0

loop_
_space_group_symop_id
_space_group_symop_operation_xyz
1 x,y,z
2 x+1/2,-y,-z
3 -x,y+1/2,-z
4 -x+1/2,-y+1/2,z
5 -y,-x,-z
6 -y+1/2,x,z
7 y,-x+1/2,z
8 y+1/2,x+1/2,-z
9 -x,-y,-z
10 -x+1/2,y,z
11 x,-y+1/2,z
12 x+1/2,y+1/2,-z
13 y,x,z
14 y+1/2,-x,-z
15 -y,x+1/2,-z
16 -y+1/2,-x+1/2,z

```
loop_
_atom_site_label
_atom_site_type_symbol
_atom_site_symmetry_multiplicity
_atom_site_Wyckoff_symbol
_atom_site_fract_x
_atom_site_fract_y
_atom_site_fract_z
_atom_site_occupancy
_atom_site_fract_symmform
Li1 Li  4 f  0.75000  0.25000  0.66794  1.00000 0,0,Dz
Li2 Li  2 c  0.25000  0.25000  0.59940  1.00000 0,0,Dz
Li3 Li  2 c  0.25000  0.25000  0.42845  1.00000 0,0,Dz
Cl1 Cl  2 c  0.25000  0.25000  0.67270  1.00000 0,0,Dz
Cl2 Cl  2 c  0.25000  0.25000  0.32256  1.00000 0,0,Dz
Cl3 Cl  4 f  0.75000  0.25000  0.58685  1.00000 0,0,Dz
Na1 Na  2 b  0.75000  0.25000  0.50000  1.00000 0,0,0
H1  H   2 c  0.25000  0.25000  0.48824  1.00000 0,0,Dz

# end of cif
```

LiCl | LiCl | NaH | LiCl | LiCl – 26%

```
# CIF file created by FINDSYM, version 7.1.3

data_findsym-output
_audit_creation_method FINDSYM

_cell_length_a    3.4925418136
_cell_length_b    3.4925418136
_cell_length_c    32.4643000000
_cell_angle_alpha 90.0000000000
_cell_angle_beta  90.0000000000
_cell_angle_gamma 90.0000000000
_cell_volume      395.9946072150

_symmetry_space_group_name_H-M "P 4/m 2/m 2/m"
_symmetry_Int_Tables_number 123
_space_group.reference_setting '123:-P 4 2'
_space_group.transform_Pp_abc a,b,c;0,0,0

loop_
_space_group_symop_id
_space_group_symop_operation_xyz
1 x,y,z
2 x,-y,-z
3 -x,y,-z
4 -x,-y,z
5 -y,-x,-z
6 -y,x,z
7 y,-x,z
8 y,x,-z
```

```
9 -x,-y,-z
10 -x,y,z
11 x,-y,z
12 x,y,-z
13 y,x,z
14 y,-x,-z
15 -y,x,-z
16 -y,-x,z

loop_
_atom_site_label
_atom_site_type_symbol
_atom_site_symmetry_multiplicity
_atom_site_Wyckoff_symbol
_atom_site_fract_x
_atom_site_fract_y
_atom_site_fract_z
_atom_site_occupancy
_atom_site_fract_symmform
Li1 Li  2 g  0.00000  0.00000  0.32323  1.00000 0,0,Dz
Li2 Li  2 h  0.50000  0.50000  0.39839  1.00000 0,0,Dz
Cl1 Cl  2 h  0.50000  0.50000  0.32324  1.00000 0,0,Dz
Cl2 Cl  2 g  0.00000  0.00000  0.41022  1.00000 0,0,Dz
Na1 Na  1 b  0.00000  0.00000  0.50000  1.00000 0,0,0
H1  H   1 d  0.50000  0.50000  0.50000  1.00000 0,0,0

# end of cif
```

## S3. Electronic structure of systems studied at null and maximum doping levels.

| | LiF \| LiH \| LiF |
|---|---|
| 0% | 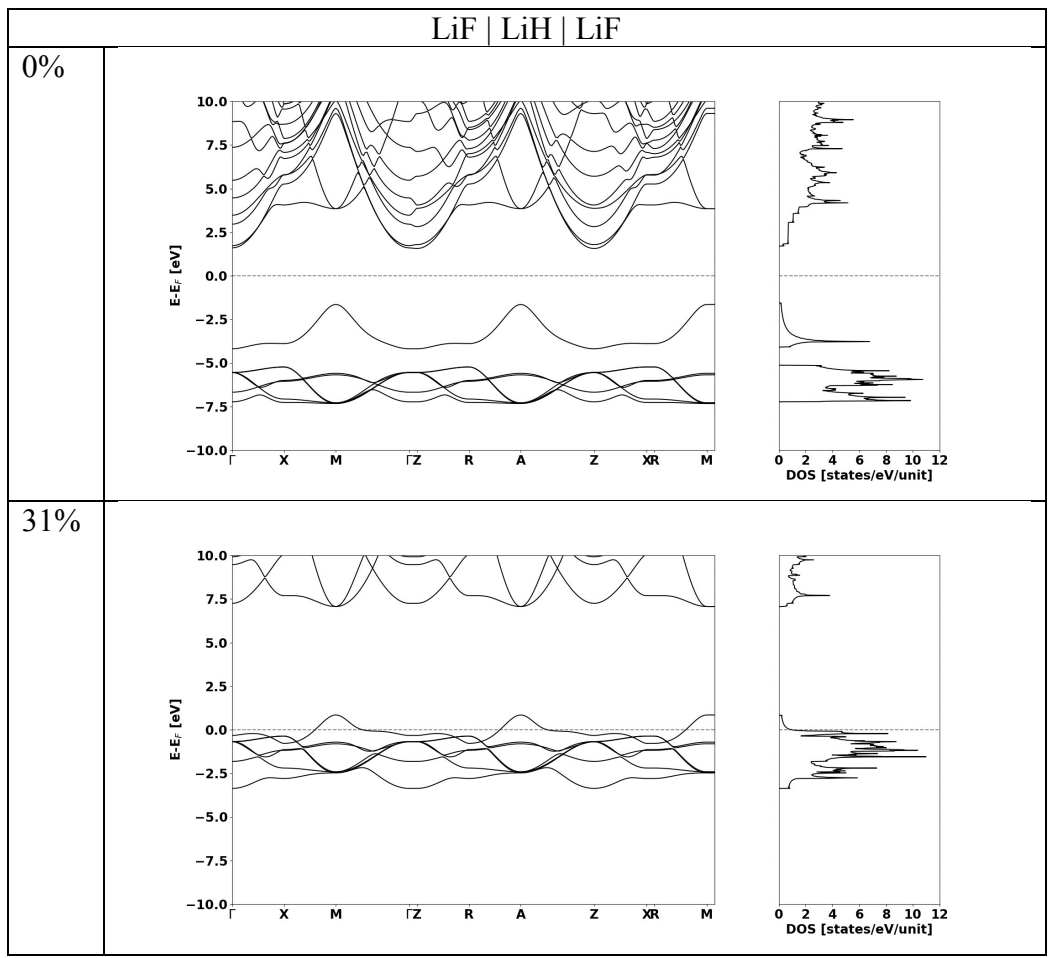 |
| 31% | |

| | LiF \| LiF \| LiH \| LiF \| LiF |
|---|---|
| 0% | 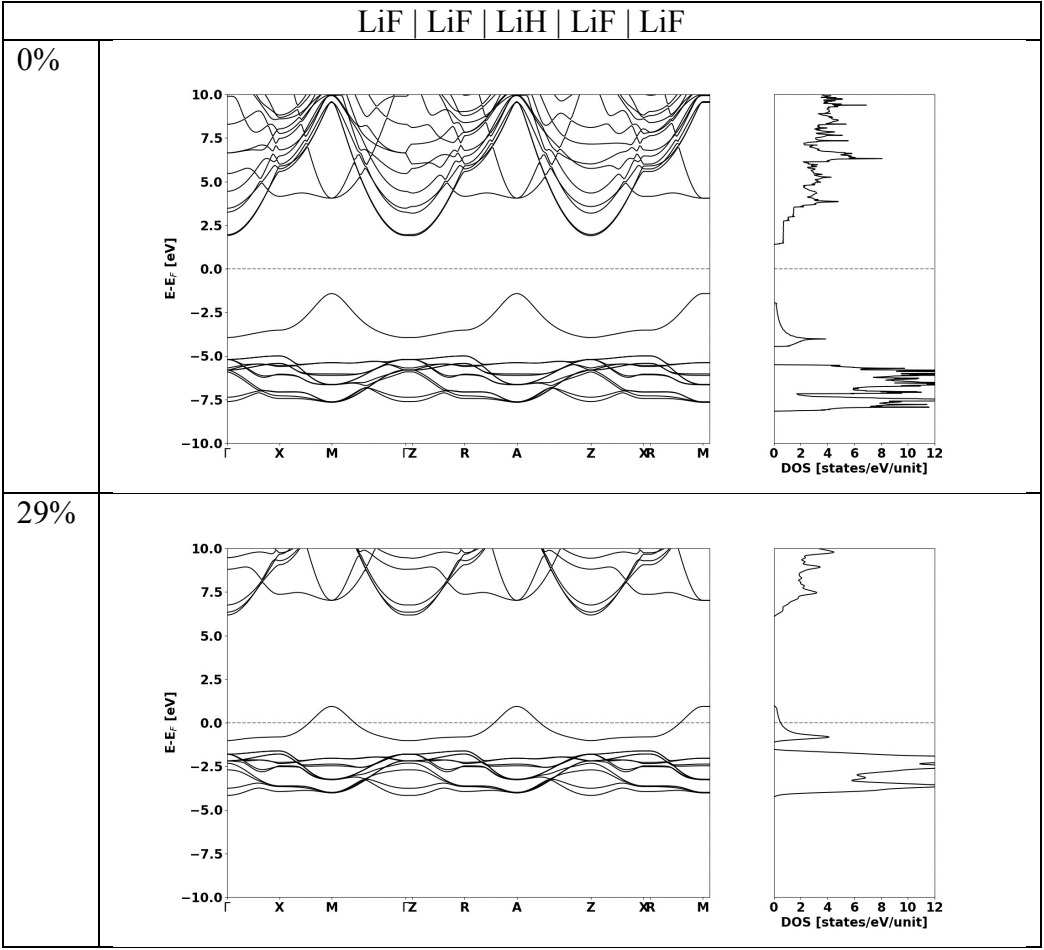 |
| 29% | |

| | KMgF$_3$ | LiH | KMgF$_3$ |
|---|---|
| 0 % | 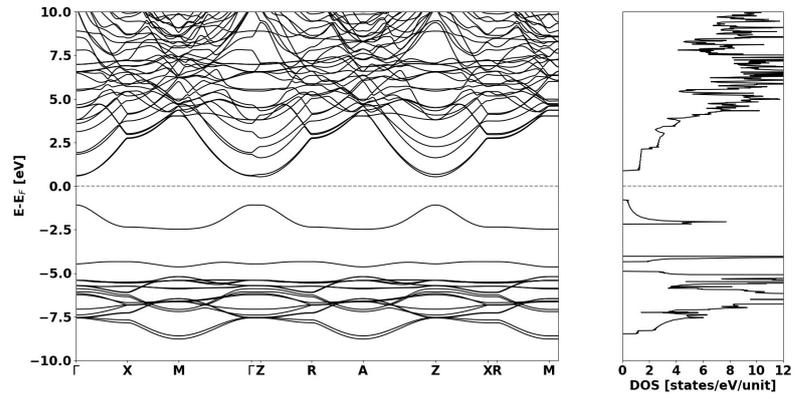 |
| 25 % | 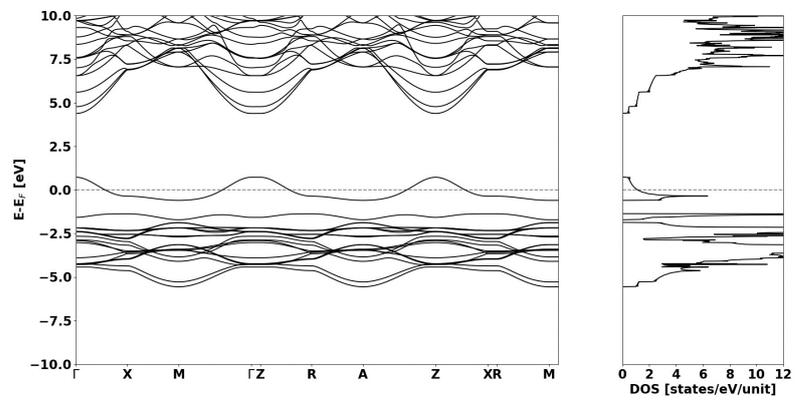 |

| | LiBaF$_3$ | LiH | LiBaF$_3$ |
|---|---|
| 0 % | 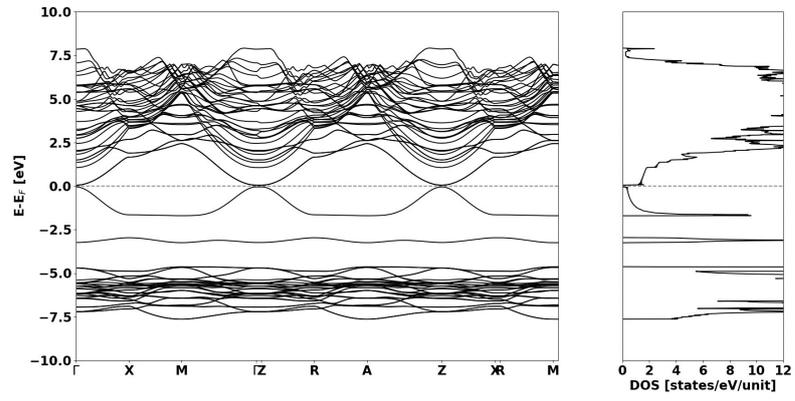 |
| 31 % | 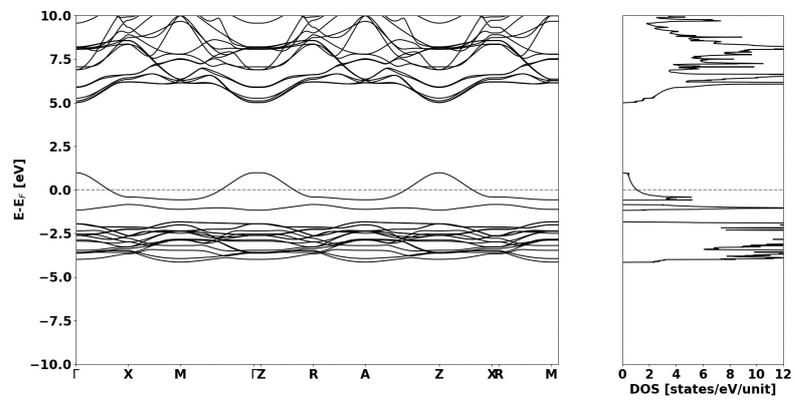 |

| | RbMgF$_3$ \| MgH$_2$ \| RbMgF$_3$ |
|---|---|
| 0% | 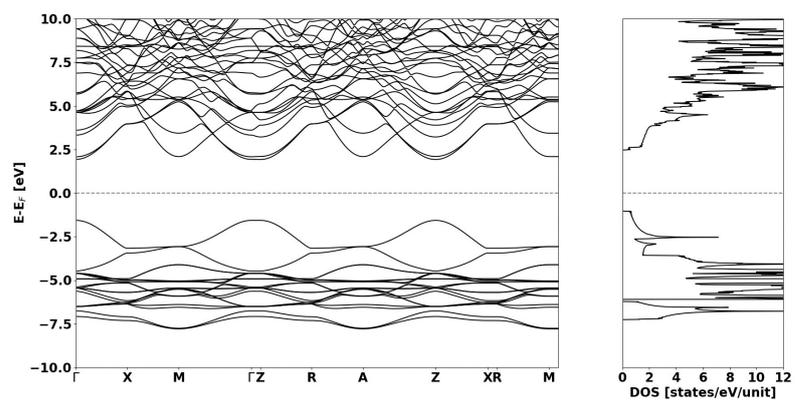 |
| 18% | 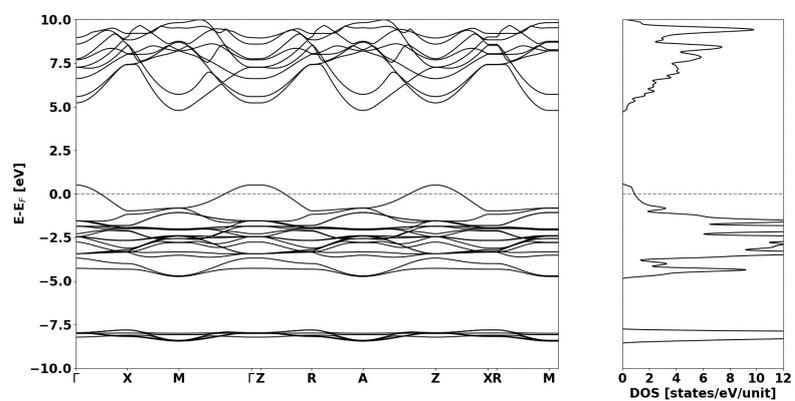 |

| LiBaF$_3$ | LiBaF$_3$ | LiH | LiBaF$_3$ | LiBaF$_3$ |

| | |
|---|---|
| 0% | 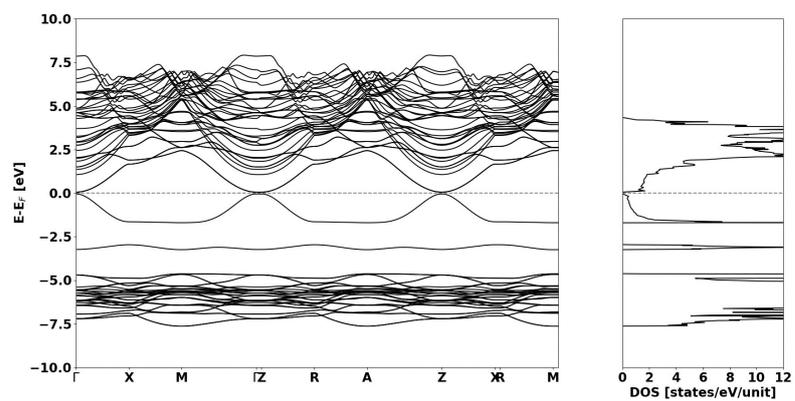 |
| 31% | 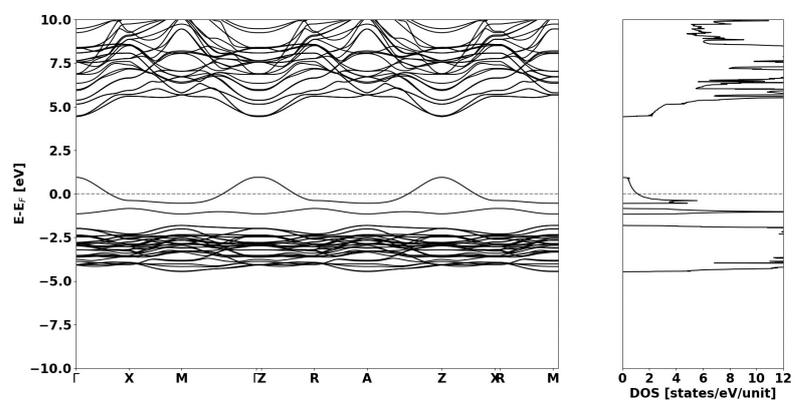 |

| | KMgF$_3$ | KMgF$_3$ | LiH | KMgF$_3$ | KMgF$_3$ |
|---|---|
| 0% | 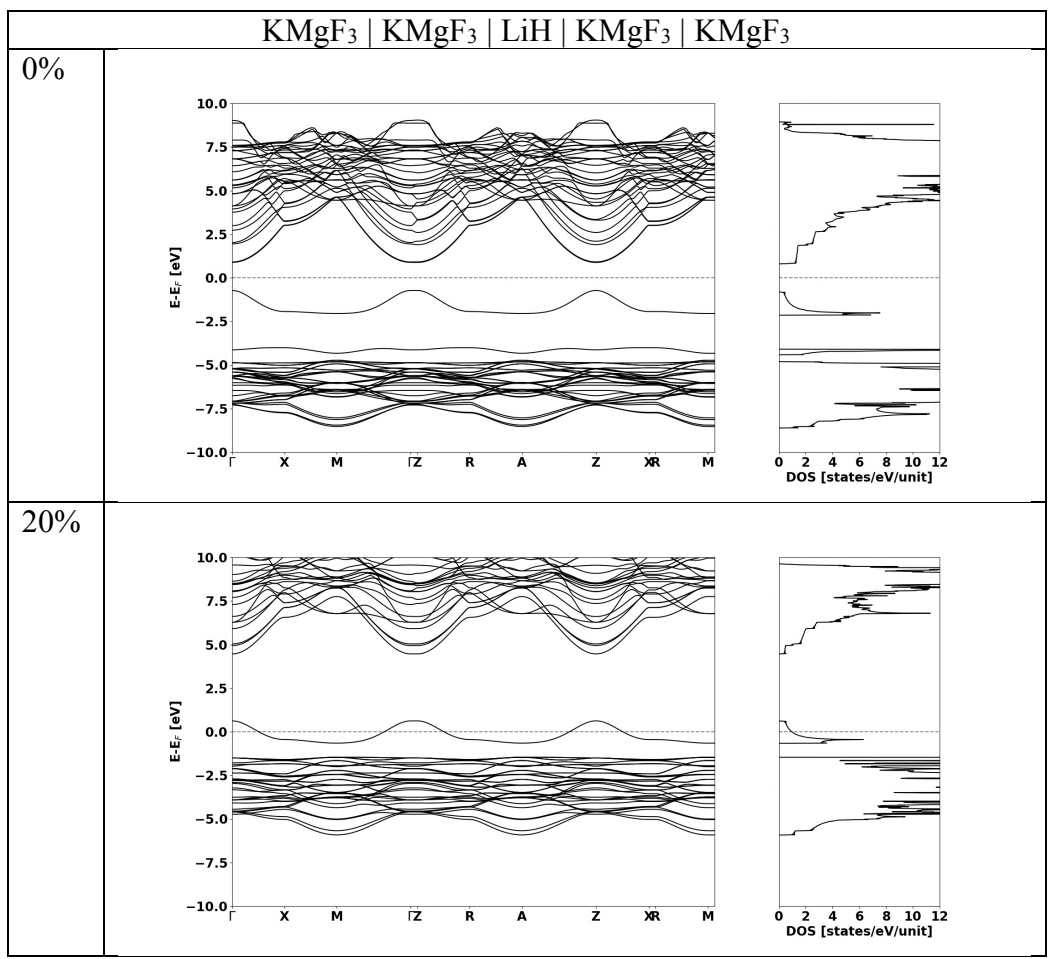 |
| 20% | |

| | RbMgF$_3$ | LiH | RbMgF$_3$ |
|---|---|
| 0% | 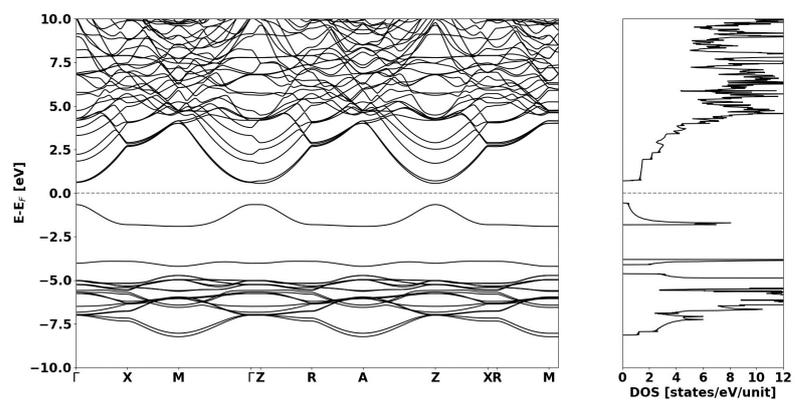 |
| 25% | 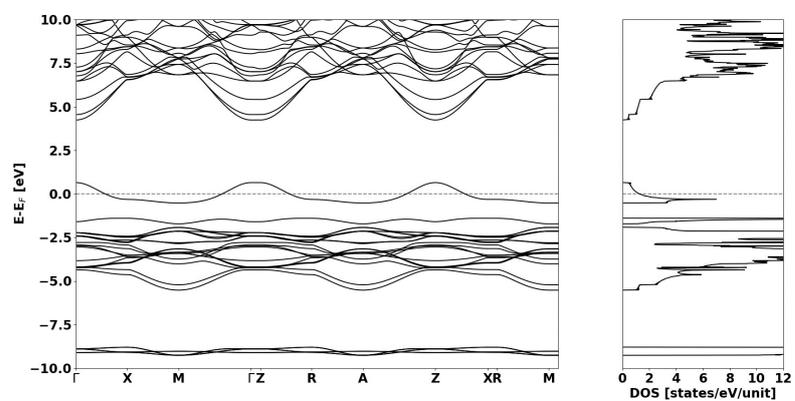 |

| | RbMgF$_3$ | RbMgF$_3$ | LiH | RbMgF$_3$ | RbMgF$_3$ |
|---|---|
| 0 % | |
| 27 % | |

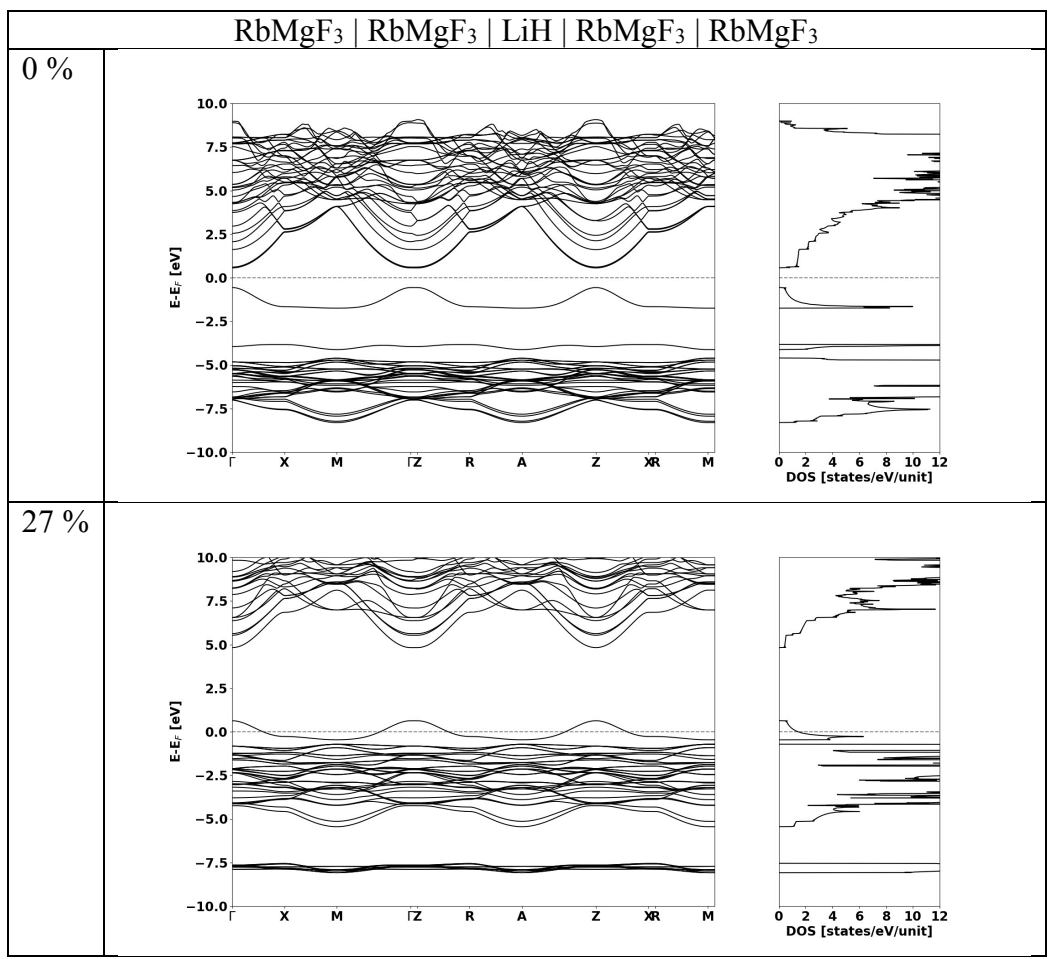

| | KZnF$_3$ | KZnF$_3$ | LiH | KZnF$_3$ | KZnF$_3$ |
|---|---|
| 0 % | |
| 5 % | |

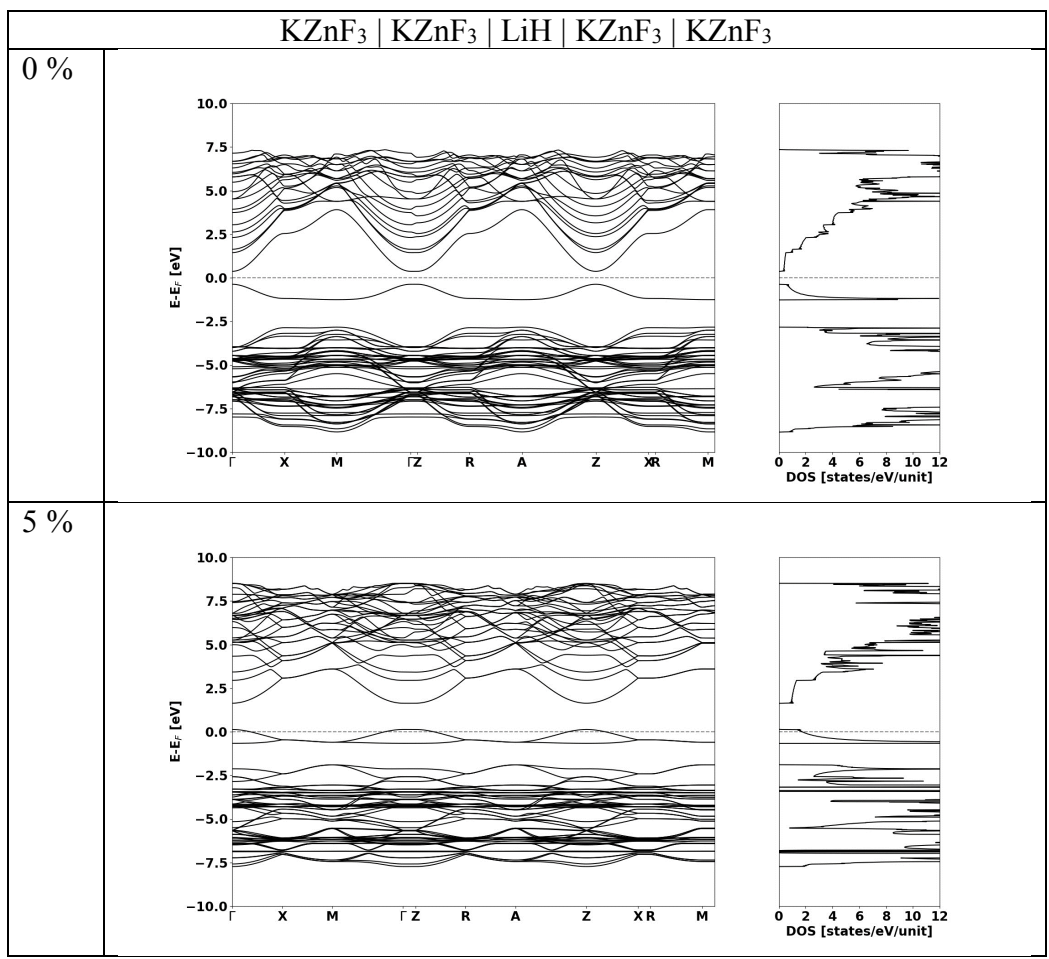

| | KZnF$_3$ | LiH | KZnF$_3$ |
|---|---|
| 0% | 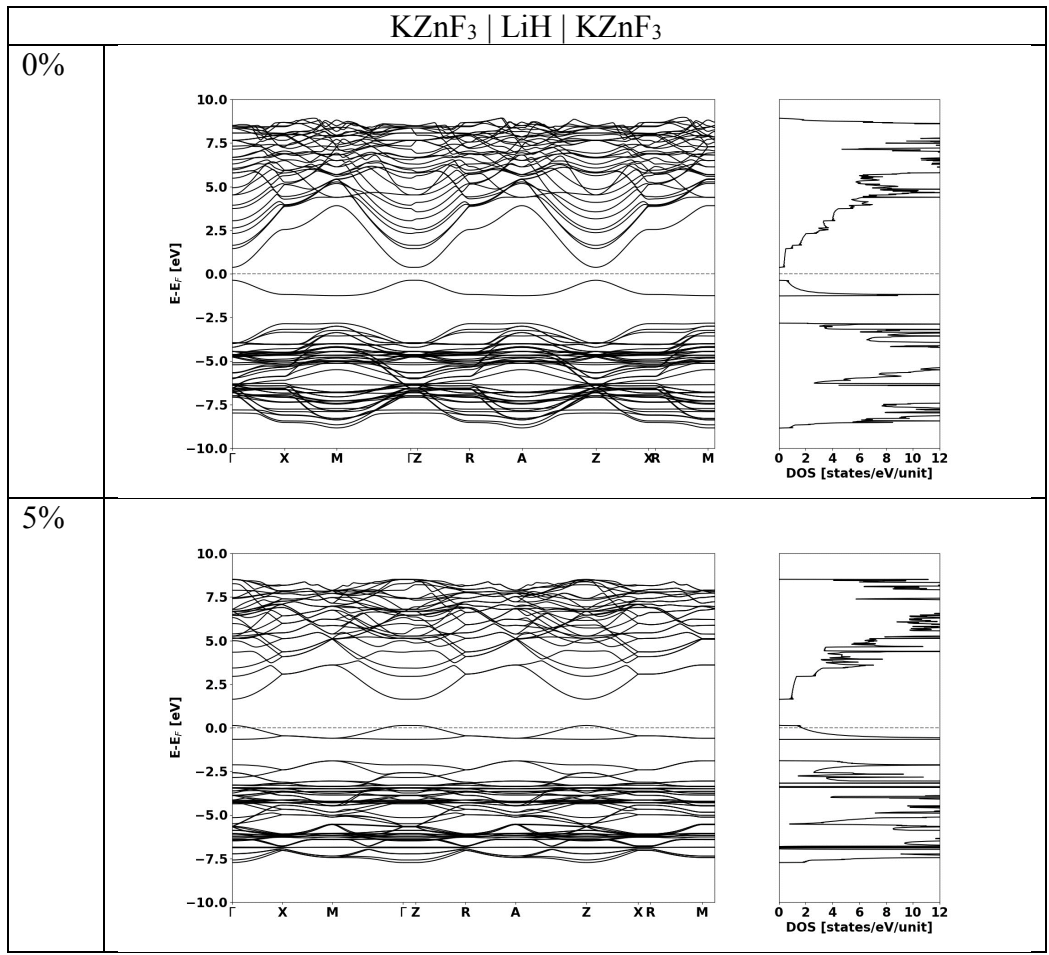 |
| 5% | |

| | CaO \| CaO \| NaH \| CaO \| CaO |
|---|---|
| 0% | 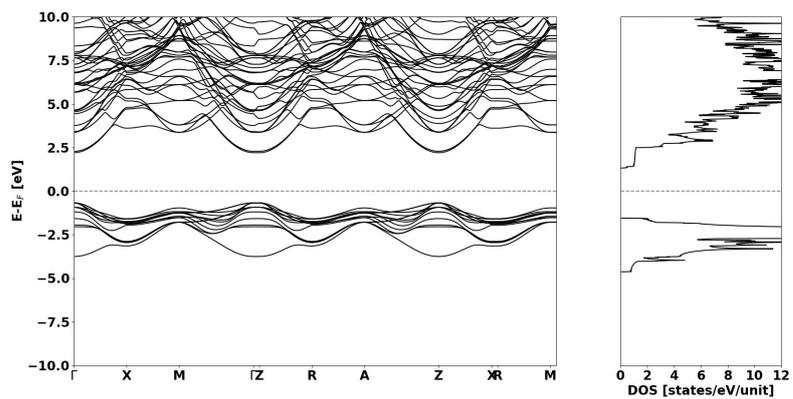 |
| 35% | 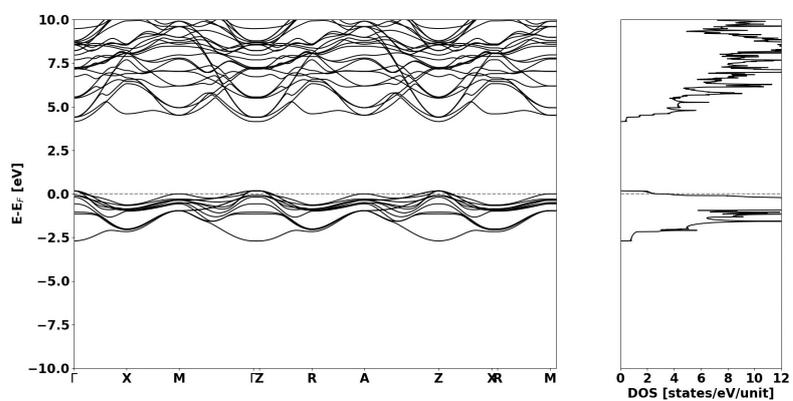 |

| | YN \| YN \| NaH \| YN \| YN |
|---|---|
| 0% | 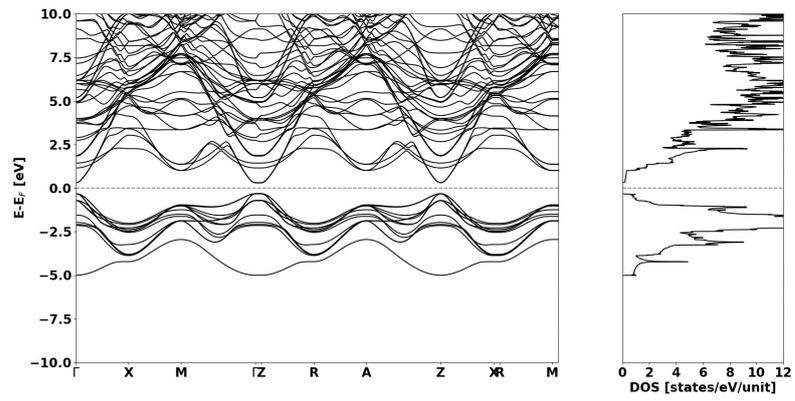 |
| 72% | 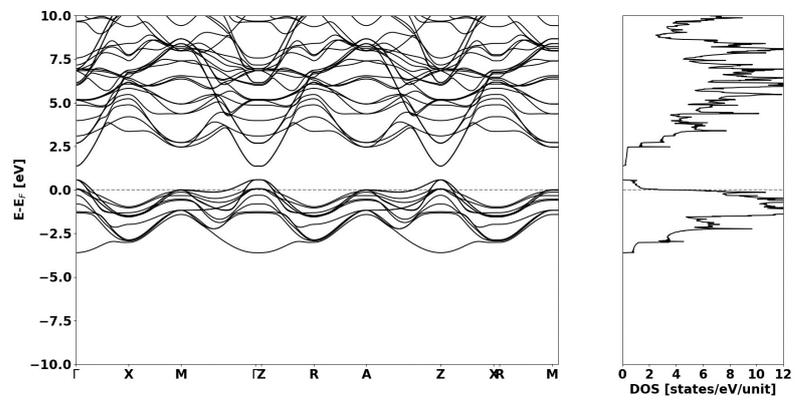 |

| | LiCl \| LiCl \| NaH \| LiCl \| LiCl |
|---|---|
| 0% | 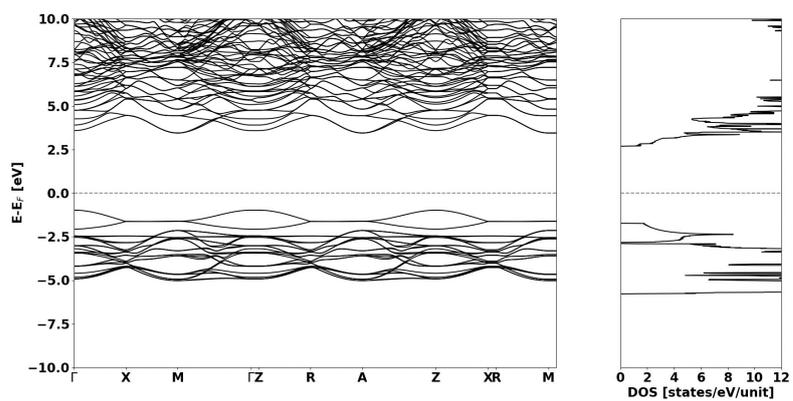 |
| 26% | 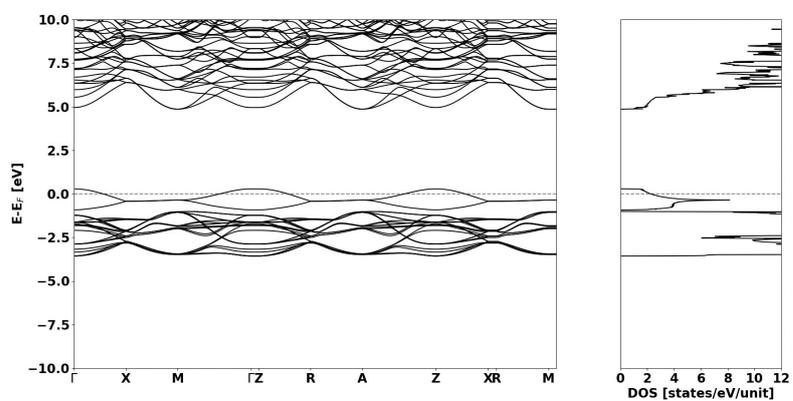 |

## S4. Phonon dispersion curves for systems studied at null and maximum doping levels.

| | LiF \| LiH \| LiF |
|---|---|
| 0% | 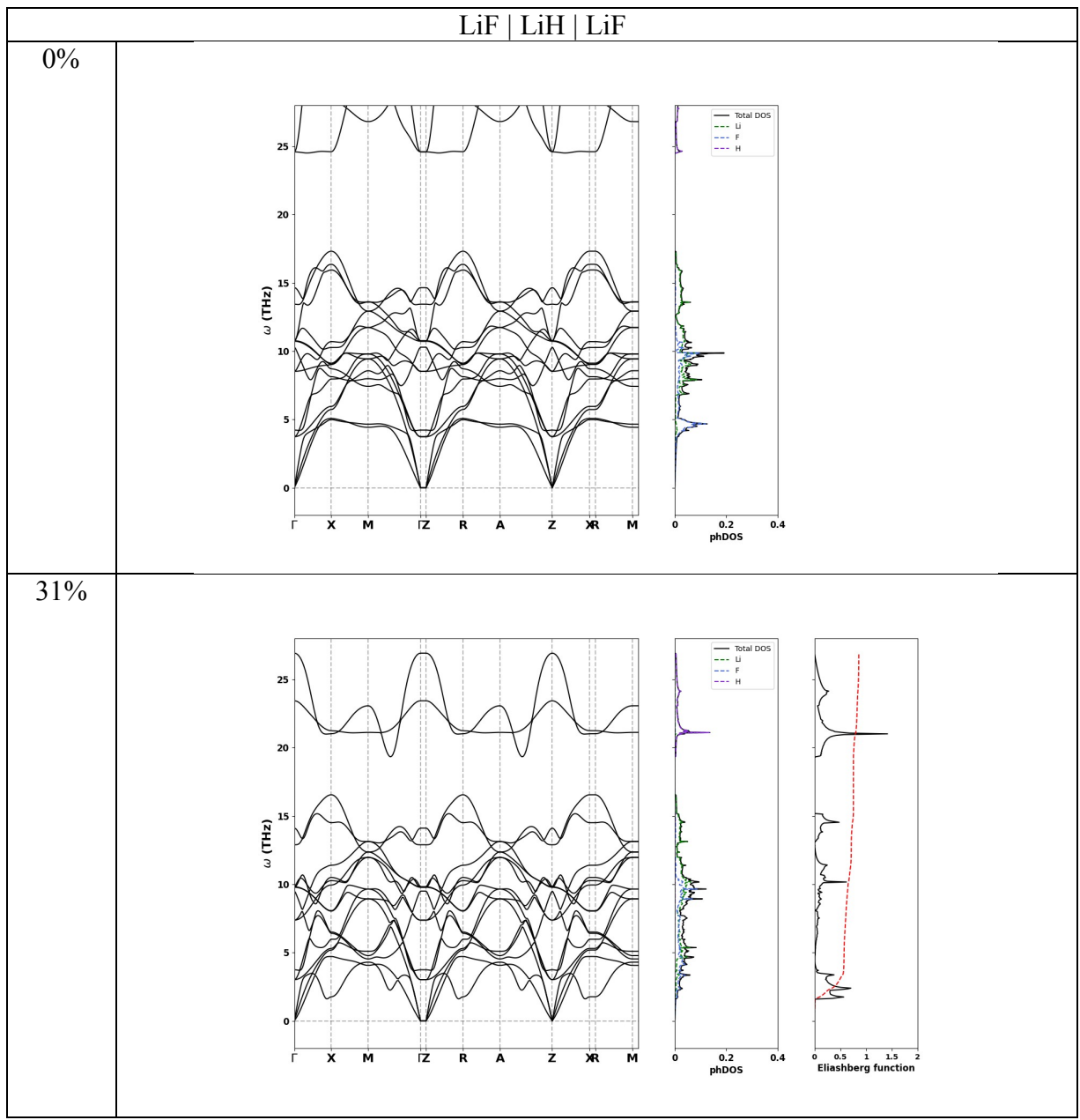 |
| 31% | |

| | LiF \| LiF \| LiH \| LiF \| LiF |
|---|---|
| 0% | 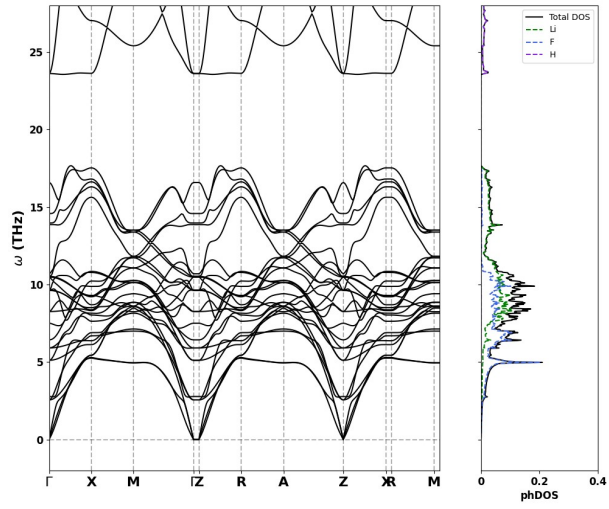 |
| 29% | 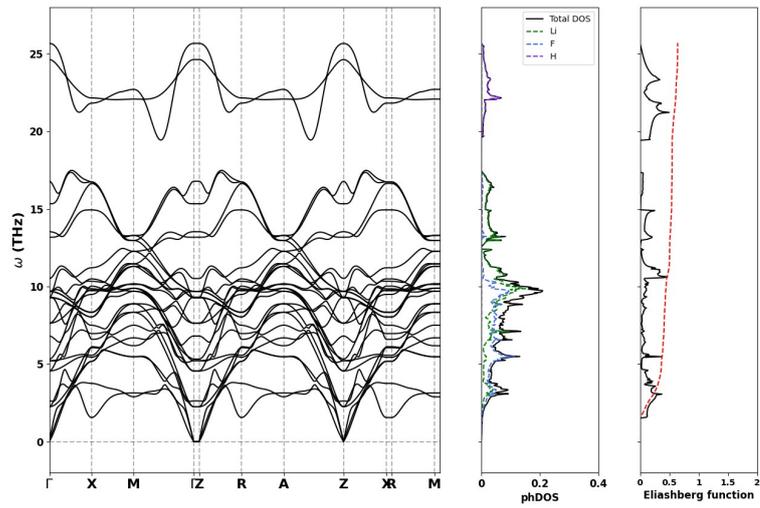 |

| | KMgF$_3$ \| LiH \| KMgF$_3$ |
|---|---|
| 0% | 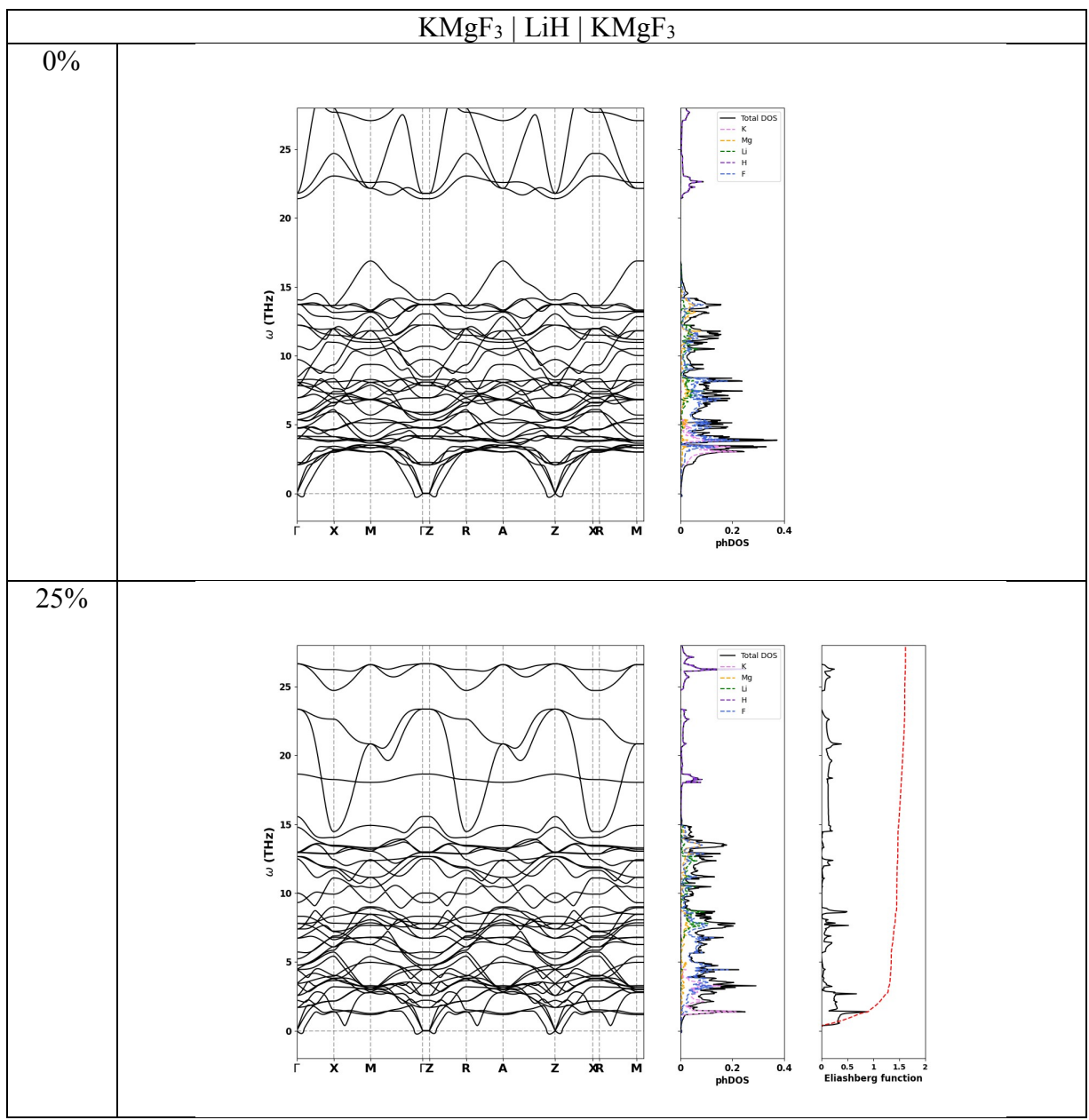 |
| 25% | |

Tiny soft features present in the vicinity of gamma and Z, but not at the exact high symmetry points, are numerical artifacts as proven by the lack of energy gain when following these normal modes.

| | LiBaF$_3$ | LiH | LiBaF$_3$ |
|---|---|
| 0% | 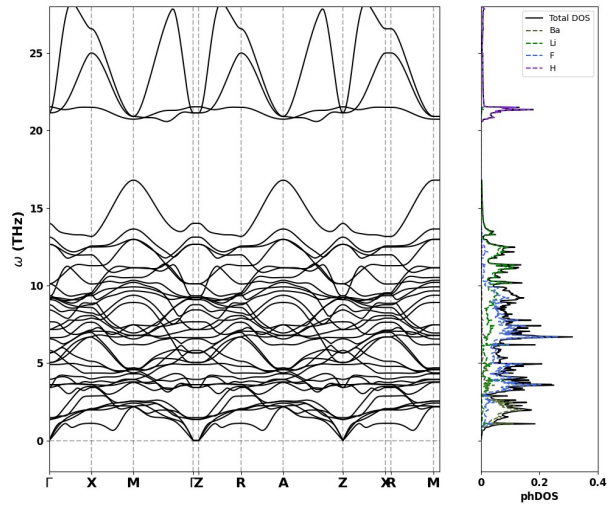 |
| 31% | 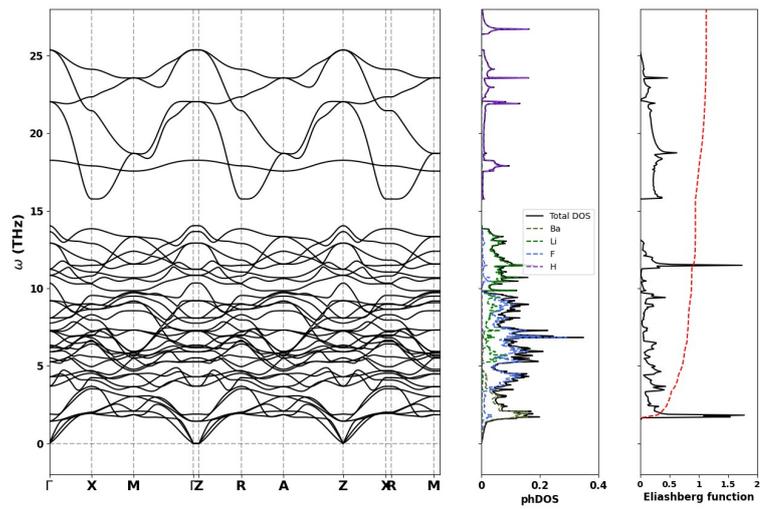 |

| | RbMgF$_3$ | MgH$_2$ | RbMgF$_3$ |
|---|---|
| 0% | 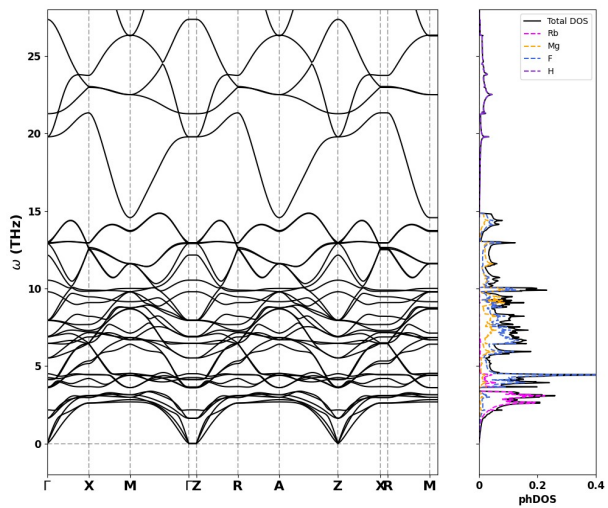 |
| 18% | 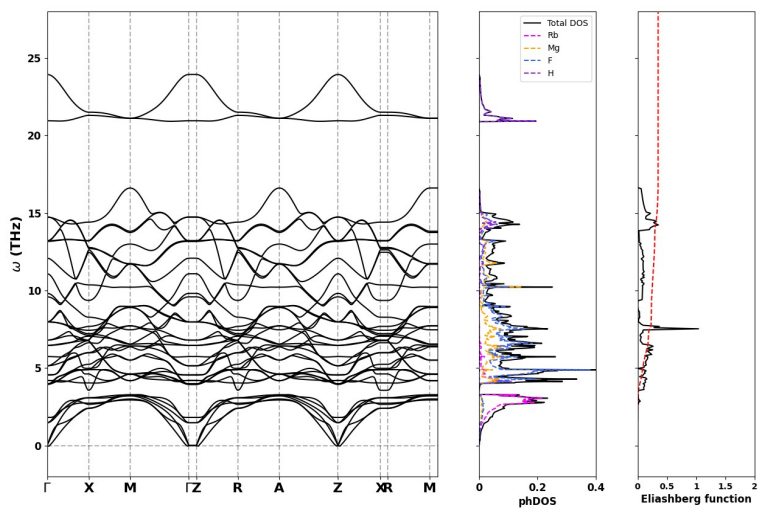 |

| | LiBaF$_3$ | LiBaF$_3$ | LiH | LiBaF$_3$ | LiBaF$_3$ |
|---|---|
| 0% | 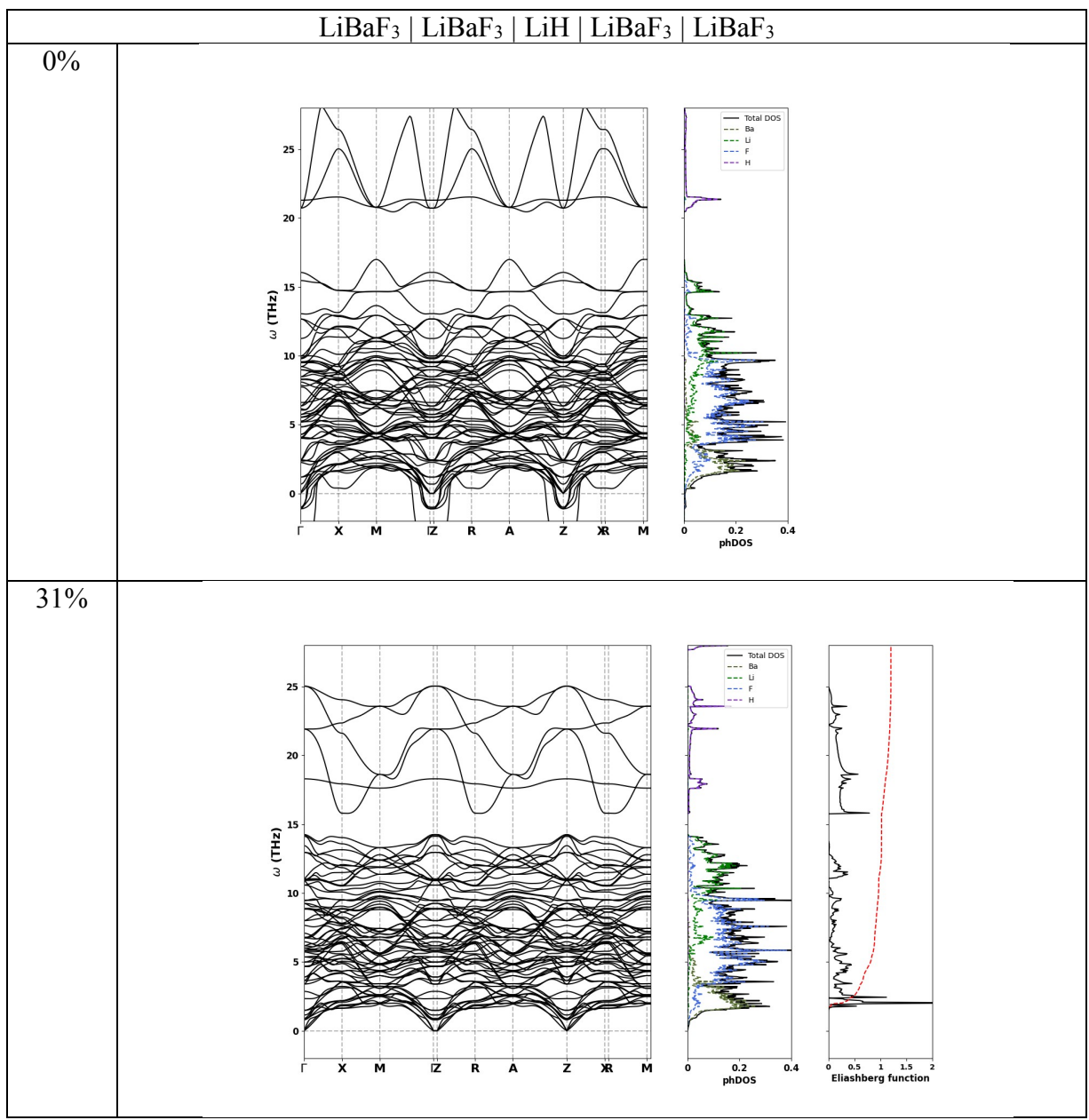 |
| 31% | |

This case represents an interesting situation where the undoped system is not dynamically stable but eventually stabilizes at an intermediate doping level of ca. 5%. Dynamical stability persists up to a maximum doping level of 31%, followed by re-entrant dynamic instability at higher doping levels.

| | KMgF$_3$ \| KMgF$_3$ \| LiH \| KMgF$_3$ \| KMgF$_3$ |
|---|---|
| 0% | |
| 20% | |

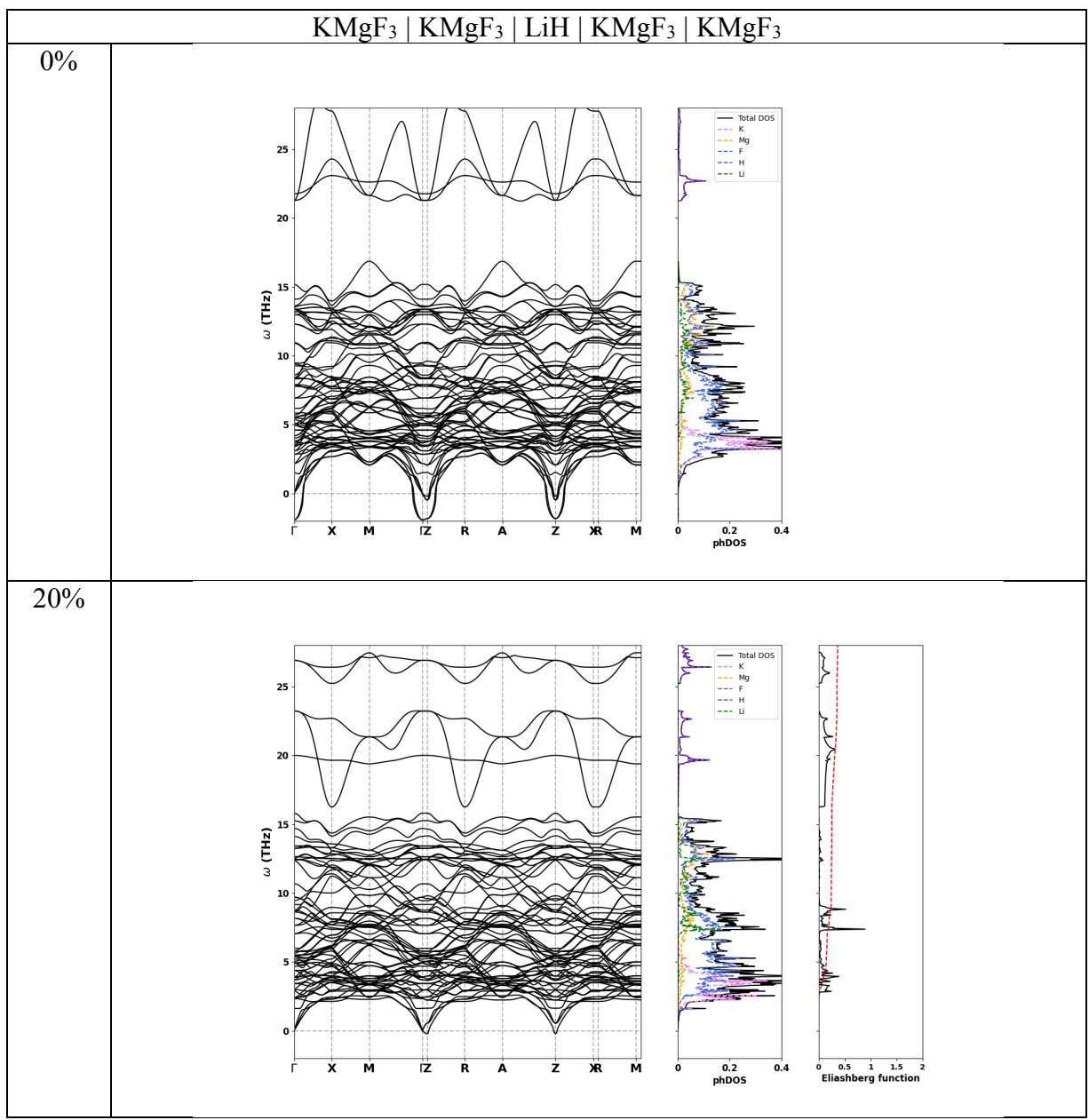

This case represents an intriguing situation where the undoped system is not dynamically stable but fully stabilizes at an intermediate doping level of ca. 5 %. Dynamic instability reappears at 15% but is dynamically stable at 20%. Further doping past 20% results in dynamic instability.

| | RbMgF$_3$ | LiH | RbMgF$_3$ |
|---|---|
| 0% | 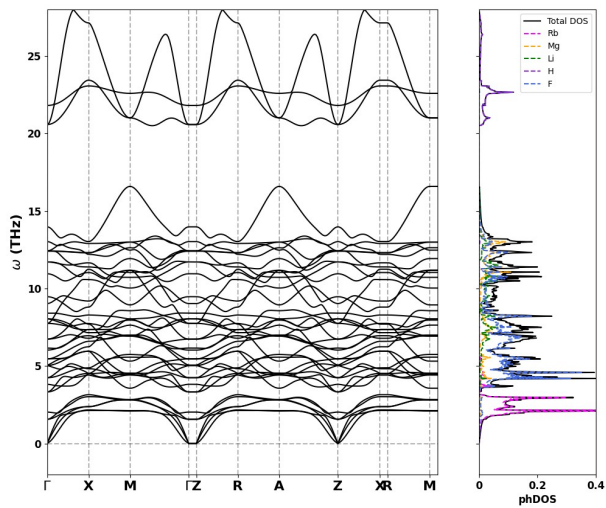 |
| 25% | 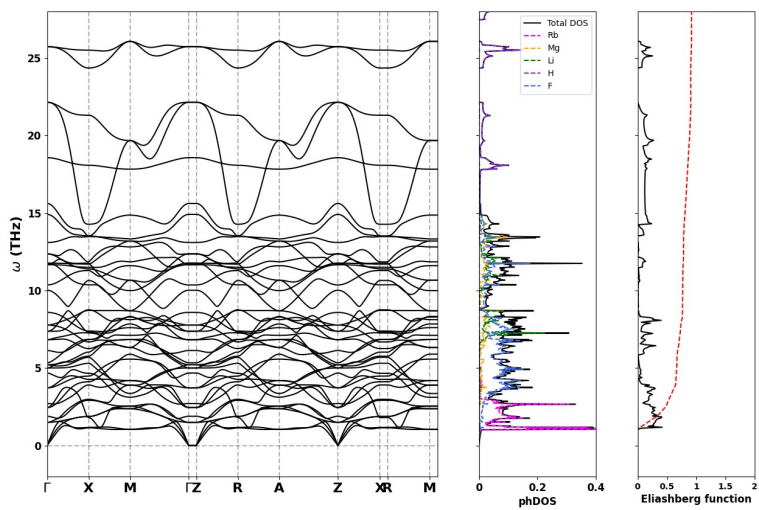 |

| | RbMgF$_3$ | RbMgF$_3$ | LiH | RbMgF$_3$ | RbMgF$_3$ |
|---|---|---|---|---|---|
| 0 % | 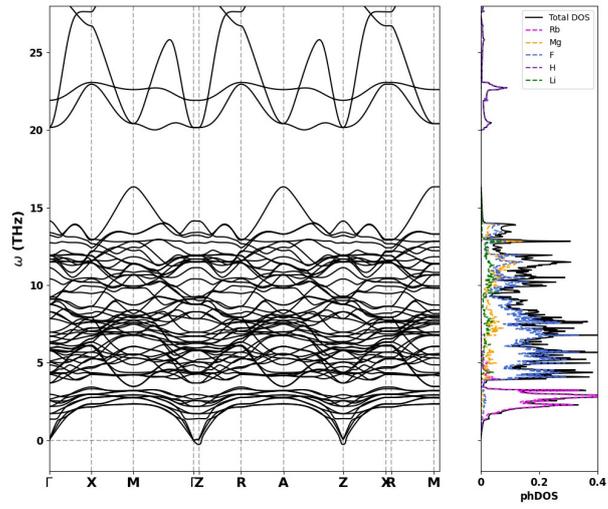 | | | | |
| 27 % | 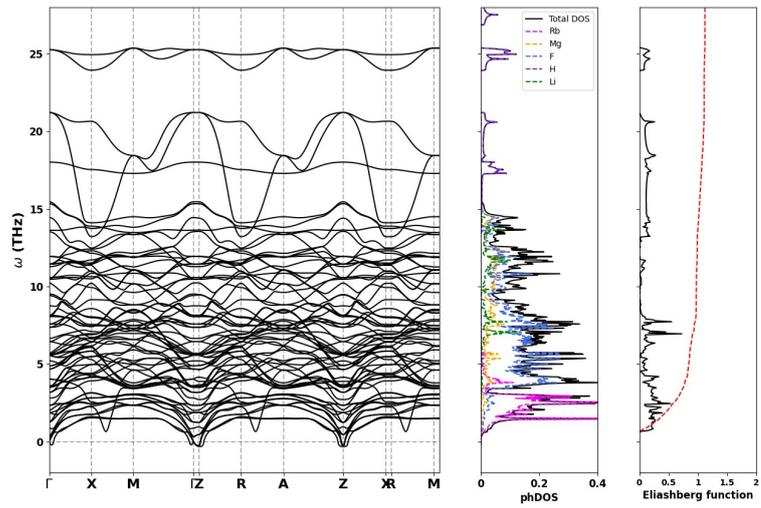 | | | | |

| | CaO \| CaO \| NaH \| CaO \| CaO |
|---|---|
| 0 % | 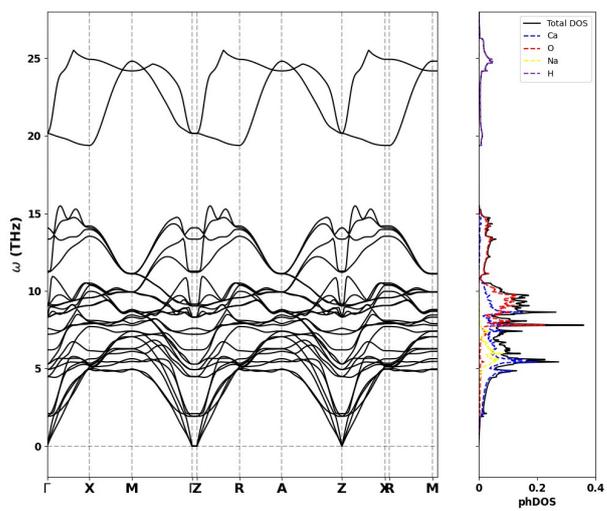 |
| 35 % | 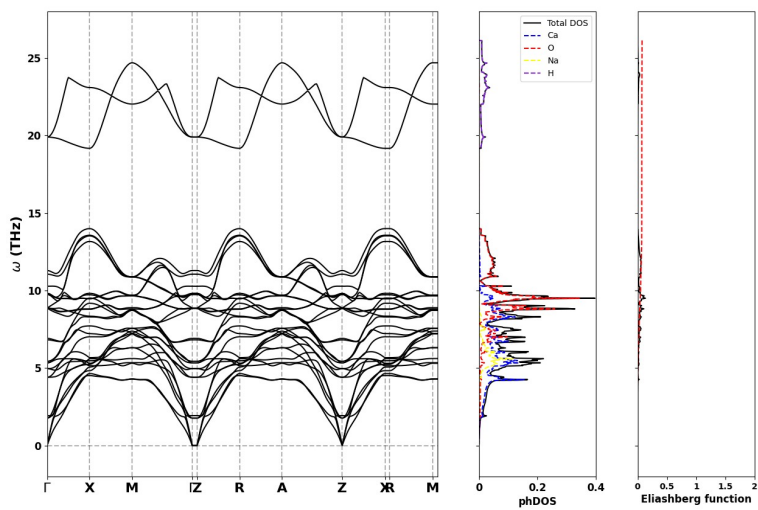 |

| | YN \| YN \| NaH \| YN \| YN |
|---|---|
| 0 % | 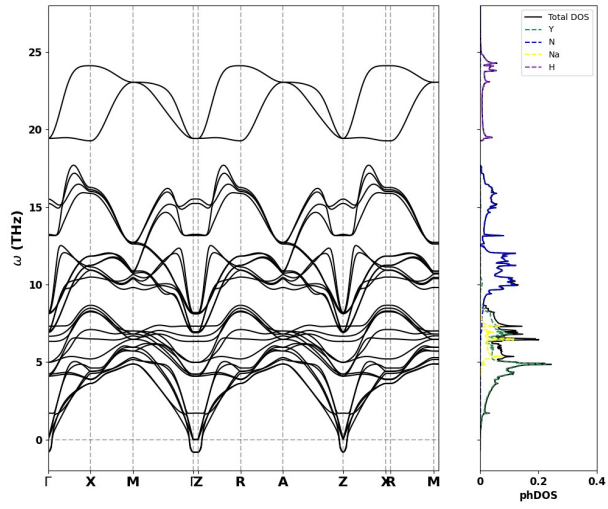 |
| 72 % | 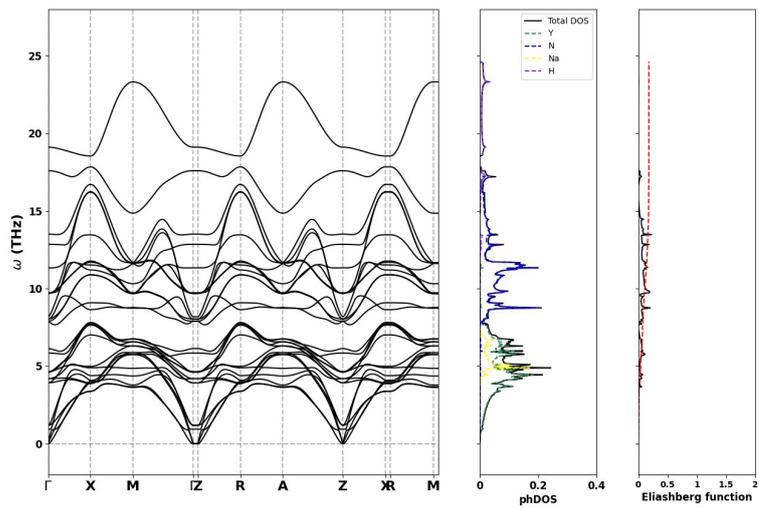 |

| | LiCl \| LiCl \| NaH \| LiCl \| LiCl |
|---|---|
| 0% | 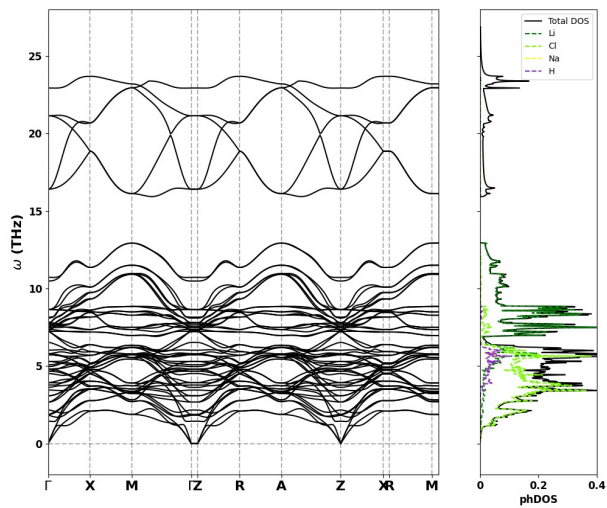 |
| 26% | 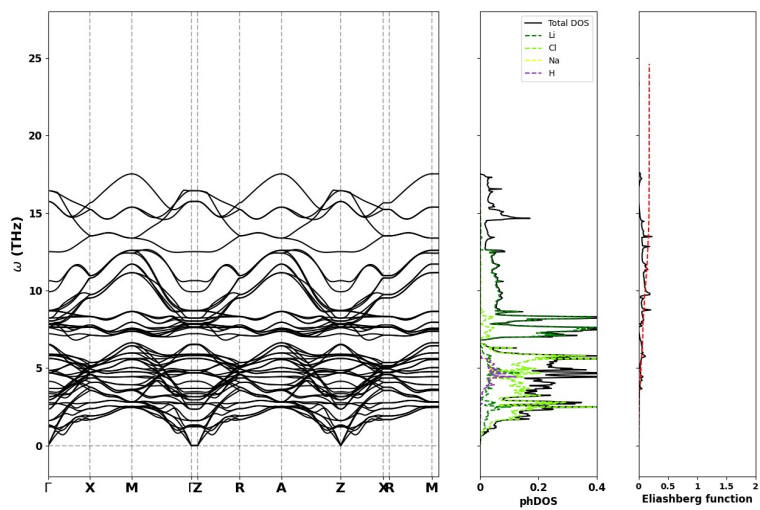 |

## S5. Dependence of DOS on a doping level as exemplified by RbMgF$_3$ | LiH | RbMgF$_3$ system.

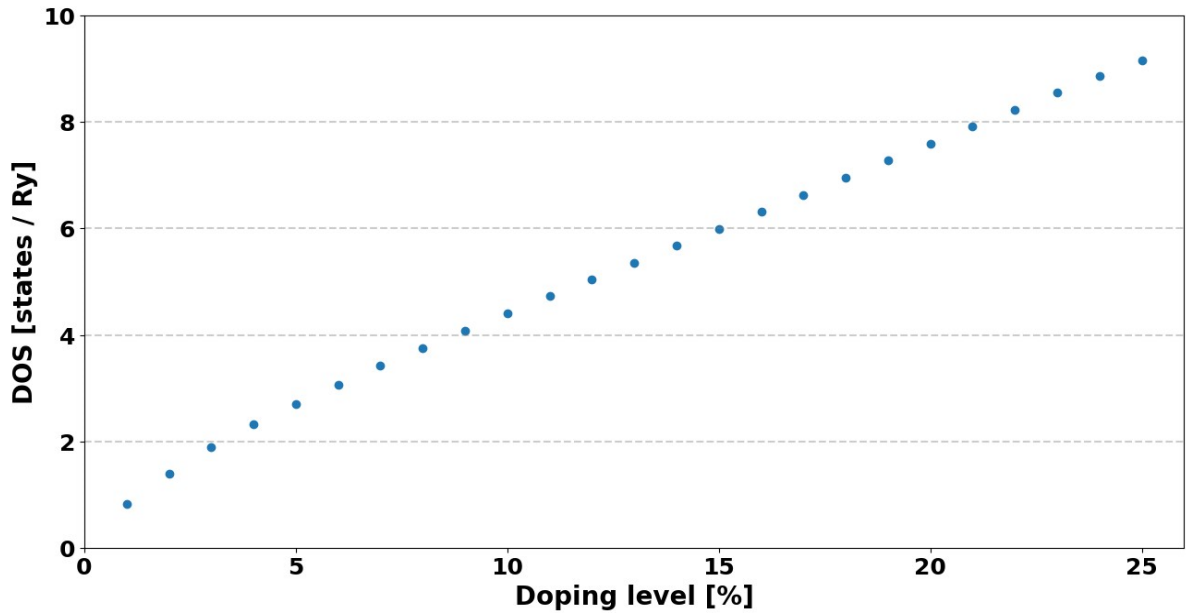

## S6. Dependence of λ on a doping level as exemplified by RbMgF$_3$ | LiH | RbMgF$_3$ system.

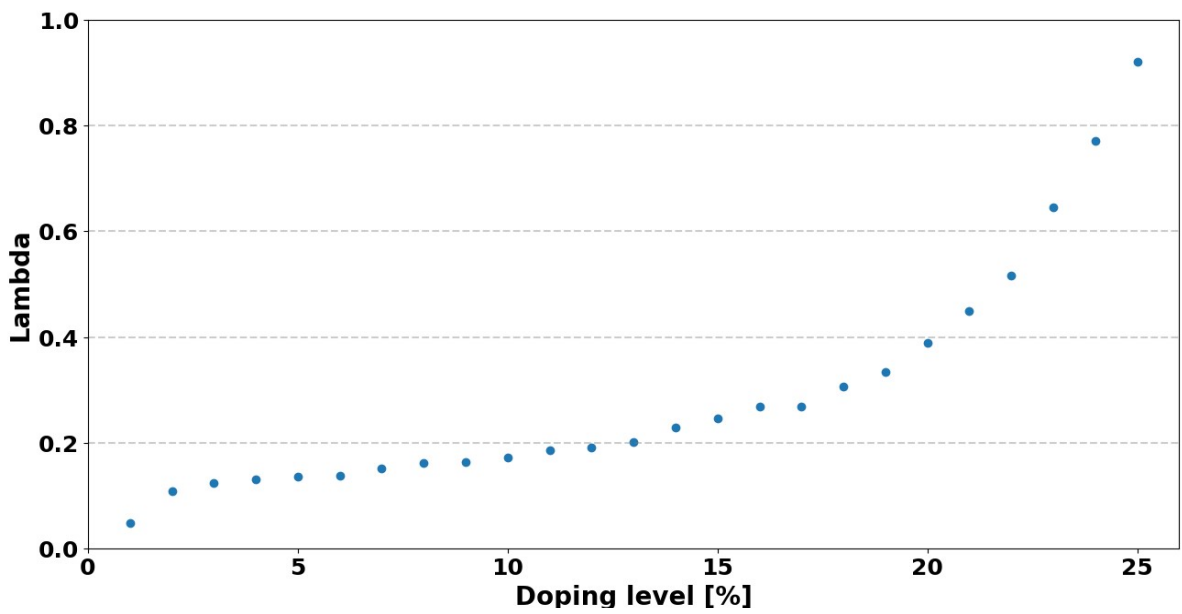

## S7. Dependence of T_C on a doping level as exemplified by RbMgF$_3$ | LiH | RbMgF$_3$ system.

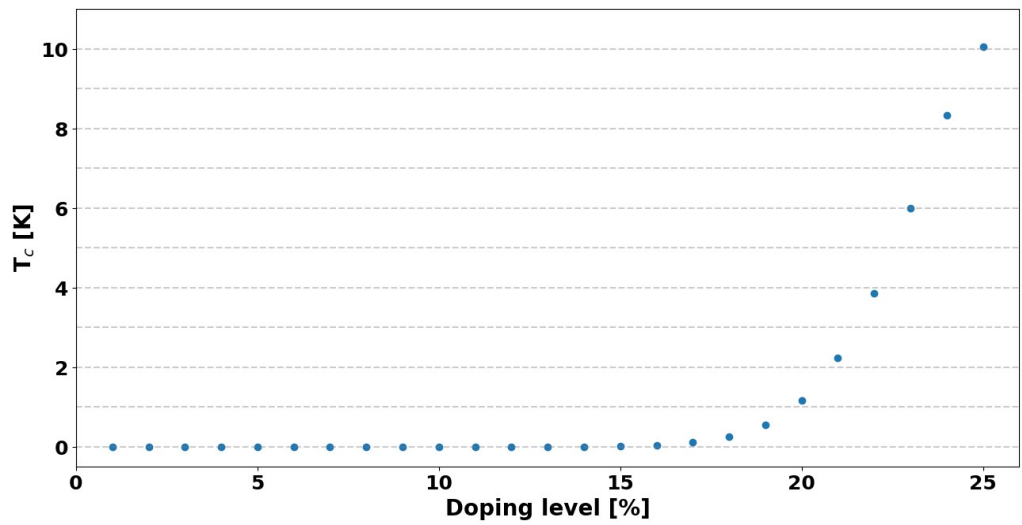

## S8. Dependence of average logarithmic frequency on doping level as exemplified by RbMgF$_3$ | LiH | RbMgF$_3$ system.

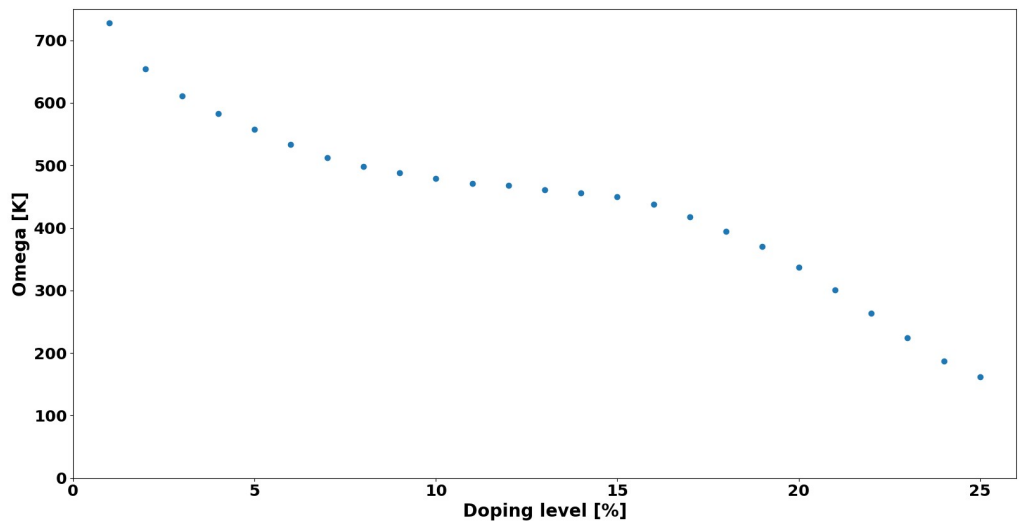

**S9. Dependence of T$_C$ on broadening parameter as exemplified by LiF | LiH | LiF system at 31% doping.**

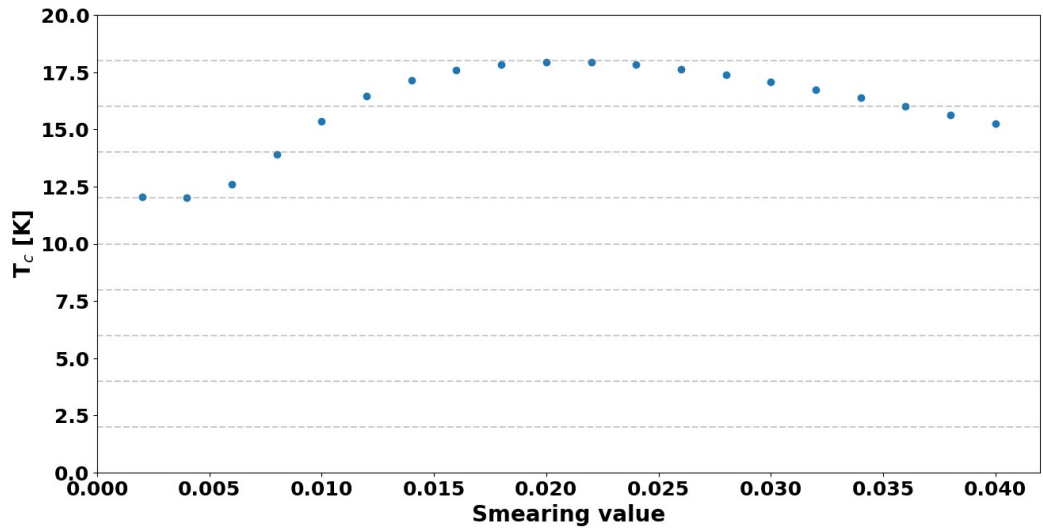

**S10. Comparison of maximum doping level as computed using QE and VASP for select systems.**

| System | Max. doping level in VASP | Max. doping level in QE |
|---|---|---|
| LiF | LiF | LiH | LiF | LiF | 31 % | 29 % |
| LiF | LiF | MgH$_2$ | LiF | LiF | 16 % | 18 % |
| (LiBaF$_3$)$_2$ | LiH | (LiBaF$_3$)$_2$ | 33.33 % | 31% |
| (RbMgF$_3$)$_2$ | LiH | (RbMgF$_3$)$_2$ | 38 % | 27 % |
| RbMgF$_3$ | LiH | RbMgF$_3$ | 27 % | 25 % |